\documentclass[12pt]{article}
\usepackage{aip}
\usepackage{amsmath}
\usepackage{graphicx}
\usepackage{amsfonts}
\usepackage{amssymb}
\begin{document}

\author{Arkady L.Kholodenko\thanks{375 H.L. Hunter Laboratories, Clemson University
,Clemson, SC 29634-1905 ,USA .e-mail: string@mail.clemson.edu}}
\title{Boundary Conformal Field Theories, Limit Sets of Kleinian Groups and Holography}
\date{The Date }
\maketitle
\begin{abstract}
In this paper, based on the available mathematical works on geometry and
topology of hyperbolic manifolds and discrete groups, some results of Freedman
et al (hep-th/9804058) are reproduced and broadly generalized.Among many new
results the possibility of extension of work of Belavin,Polyakov and
Zamolodchikov to higher dimensions is investigated. Known in physical
literature objections against such extension are removed and the possibility
of an extension is convinsingly demonstrated
\end{abstract}

{\LARGE 1. Introduction}

\bigskip

Recently, there had been attempts to extend the results of two dimensional
conformal field theories(CFT) to higher dimensions $[1.2].$ Since publication
of papers by Witten$[3,4]$it had become clear that there is a very close
correspondence between 2d physics of critical phenomena and 3d physics of
knots and links. A very detailed study of this correspondence is developed by
Moore and Seiberg $[5]$. Additional contributions more recently were made in
Ref[6], etc. All these works heavily exploit the algebraic aspects of this
correspondence through use of Yang-Baxter equations, quantum groups, etc. Much
lesser efforts had been spent on development of the same correspondence from
the topological point of view through study of 3-manifolds complementary to
knots(links) in $S^{3}$=$R^{3}\cup\{\infty\}.$ Such study is potentially more
beneficial since it is known $[7]$ that in four dimensions \textbf{all} knots
are trivial (i.e.unknotted) so that the algebraic methods used so far are
necessarily limited to 3 dimensions and, accordingly, to study of two
dimensional CFT only. At the same time, topological study of manifolds is not
limited to three dimensions. The reasons why such studies are useful could be
understood from the following simple arguments taken from the book by Maskit $[8].$

Define an inclusion of $R^{d}$ into $R^{d+1}$ through $R^{d}$ =\{(x,t)$\mid
t=0\}$ where x$\in R^{d},-\infty\leq t\leq\infty.$ The upper half space
Poincar$e^{\prime}$ model of hyperbolic space H$^{d+1}$is defined by
\begin{equation}
\text{H}^{d+1}=\{(x,t)\mid t>0\} \tag{1.1}%
\end{equation}
with $x\in R^{d}$ so that $\partial$H$^{d+1}=R^{d}$. Consider a special group
G of motions G=M$^{d+1}$ of $R^{d+1}=\{x,t\}$ made of

a) translations: \quad(x,t)$\rightarrow(x+a,t)$ ,$a\in R^{d-1\text{ }};$

b) rotations :\quad\quad(x,t)$\rightarrow$ (r(x),t) ,\quad r$\in O(d-1);$

c) dilatations :\quad\quad x$\rightarrow\lambda x$ ,$\lambda>0,\lambda\neq1$ and

d) inversions :\quad\quad x$\rightarrow\frac{x}{\left|  x\right|  ^{2}}$ .
\quad\quad\quad\quad\quad\quad\quad\quad\quad\quad\quad\quad\quad\quad
\quad\quad\quad\quad\quad\quad\quad\quad\quad

It can be proven$[8]$ that the group G acts as a group of \textbf{isometries
}of H$^{d+1}$and is called d dimensional M$\ddot{o}$bius group. In its action
on R$^{d}$ ''G acts as a group of conformal motions but \textbf{not }as a
group of isometries in \textbf{any} metric''.

At the same time, it is well established$[9]$ that in \textbf{any dimension}
the physical system at criticality possesses the invariance which is described
in terms of the group G. Hence, \textbf{the very existence of}
\textbf{criticality is closely associated with the hyperbolicity of the
adjacent space.}

Let x$\in H^{d+1}$ and $\gamma\in$G. Consider a motion (an orbit) in H$^{d+1}$
by successive applications of $\gamma$ to x. It is of interest to study if
such a motion will ever hit $\partial$H$^{d+1}=R^{d}$. This problem is highly
nontrivial and was solved by Beardon and Maskit$[10]$ (e.g.see section 5 below
for more details) for d=2. The nontriviality of this problem could be better
understood if, instead of the upper half space H$^{d+1}$model, we would
consider the unit ball B$^{d+1}$ model of the hyperbolic space with the unit
sphere $S_{\infty}^{d}$ $($sphere at infinity) playing the same role as in
this model as $\partial$H$^{d+1}=R^{d}$ in the upper half space model. Since
not all subgroups of G will allow hitting of the boundary, it is clear, that
one should be interested only in those subgroups whose orbits end up at the
boundary. These subgroups, in turn, could be further subdivided into those
whose limit points on $S_{\infty}^{d}$ will cover the entire sphere and those
which will cover only a part of $S_{\infty}^{d}$. This part we shall denote as
$\Lambda.$ The limit set $\Lambda$ is actually a fractal . The fractal
dimension of $\Lambda$ is directly related to the critical indices of the
two-point correlation functions of the corresponding conformal models at
criticality. Different subgroups of M$\ddot{o}$bius group G will produce
different fractal dimensions. In turn, the corresponding hyperbolic manifolds
associated with these groups could be viewed as complements of the related
knots (links) in the case of 2+1 dimensions so that different conformal
models, indeed, could be associated with different types of knots(links). This
association becomes unnecessary when one is interested in conformal models in
dimensions 3 and higher. One could still consider motions associated with
subgroups of M$\ddot{o}$bius group and the corresponding, say, hyperbolic
4-manifolds without using knots, braids, Yang-Baxter equations, etc.

Although stated in different form, recent results of Maldacena$[11]$ and their
subsequent refinements in Refs[12-16] (and many additional references therein
and elsewhere which we do not include) are actually directly connected with
ideas just described. In physics literature the connection between ''surface''
and ''bulk'' field theories is known as \textbf{holographic}
\textbf{principle} (holographic hypothesis)[17,18]. In simple terms [19], it
can be formulated as statement that ''a macroscopic region of space and
everything inside it can be represented by a boundary theory living on the
boundary region''. Mathematical support of this principle in physics
literature is attributed to works by Fefferman and Graham [20] and Graham and
Lee[21]. These works discuss \ boundary conditions at infinity for Einstein
manifolds (spaces) and initial value problem for Einstein's equations.
Although our previous discussion did not involve the Einstein manifolds,
actually, the results of Ref.[21] are consistent with those which follow from
the hyperbolic geometry. This can be understood if one takes into account that
Einstein spaces are characterized by the property that the Ricci tensor
$R_{ij}$ is proportional to the metric tensor $g_{ij}$ [22], that is
\begin{equation}
R_{ij}=\lambda g_{ij}.\tag{1.1}%
\end{equation}
Since the scalar curvature $R=g^{ij}R_{ij}$ , the above equation can be
rewritten as%
\begin{equation}
R_{ij}=\frac{R}{d}g_{ij}\tag{1.2}%
\end{equation}
where d is the dimensionality of space (as before). The Einstein tensor
\begin{equation}
G_{j}^{i}=R_{j}^{i}-\frac{1}{2}\delta_{j}^{i}R\tag{1.3}%
\end{equation}
acquires a particularly simple form with help of Eq.(1.2):%
\begin{equation}
G_{j}^{i}=(\frac{1}{d}-\frac{1}{2})\delta_{j}^{i}R\tag{1.4}%
\end{equation}
and, because $G_{j,h}^{i}$ =$0$, we obtain,%
\begin{equation}
(\frac{1}{d}-\frac{1}{2})R\text{ ,}_{j}=0.\tag{1.5}%
\end{equation}
This implies ,that the scalar curvature $R$ is constant . For isotropic
homogenous spaces E$_{d}$ the Riemann curvature tensor is known to be[23]
given by
\begin{equation}
R_{ijkl}=k(x)(g_{ik}g_{jl}-g_{il}g_{jk})\tag{1.6}%
\end{equation}
so that the Ricci tensor \ is \ given by
\begin{equation}
R_{ij}=(d-1)k(x)g_{ij},\tag{1.7}%
\end{equation}
where k(x) is the sectional curvature at \ the point x$\in$E$_{d}$ . The
Schur's theorem[23] guarantees, that $k(x)=k=const$ for $d\geq3$ .Comparison
between Eqs.(1.1) and (1.7) produces then : $\lambda=(d-1)k$ and, accordingly,
$R=d(d-1)k$ . The spatial coordinates can always be rescaled so that, for
$k<0$ we obtain, the canonical value $k=-1$ characteristic of hyperbolic
space[24,25] . Since \ in the work by Graham and Lee [21] the condition given
by Eq.(1.7) is used (with $k=-1$), the connections with hyperbolic geometry
\ is evedent. Since Eq.(1.4) can be equivalently rewritten with help of
Eq.(1.7) as
\begin{equation}
R_{ij}-\frac{1}{2}g_{ij}R+\hat{\Lambda}g_{ij}=0,\tag{1.8}%
\end{equation}
with the cosmological term $\hat{\Lambda}$ =-$\frac{1}{2}(d-1)(d-2),$thus
obtained equation produces metric for Einstein space \ known in literature as
anti-de Sitter (AdS) space[26]. Hence, in part, the purpose of this work is to
investigate in some detail connections between the results obtained in physics
literature and related to CFT-AdS correspondence, e.g.see Ref.[12], and those
known in mathematics those known in mathematics and related to hyperbolic
geometry and hyperbolic spaces. Not only it is possible to reobtain results
known in physics using these connections, but many more follow along the way
of physical reinterpretation of known results in mathematics. Establishing
these connections touches many aspects of modern mathematics such as the
geometry and topology of hyperbolic manifolds$[25]$, multidimensional
extension of the theory of Teichm\"{u}ller spaces$[27],$ spectral analysis of
hyperbolic manifolds$[28]$ (including those with cusps$[29])$, random walks on
group manifolds$[30]$, theory of deformations of Kleinian and Fuchsian
groups$[31]$(and M$\ddot{o}$bius groups in general), ergodic theory of
discrete groups$[32]$, Kodaira-Spencer theory of deformations of complex
manifolds$[33]$, loop groups$[34],$cohomology of groups, etc. In particular,
the cohomological aspects of these connections lead directly to the Virasoro
algebra and its generalizations thus allowing us to dicuss the extension of
fundamental results of Belavin-Polyakov-Zamolodchikov(BPZ) [35] to higher
dimensions(e.g.see section 8). To make our presentation self-contained, we had
incorporated some the auxiliary results from mathematics into text which are
meant only to facilitate reader's understanding without detracting of his/her
attention from \textbf{physical} goals and motivations of this work. A quick
summary of some auxiliary mathematical results related to hyperbolic
3-manifolds and Einstein spaces also could be found in our papers, Ref.s[36,37].

This paper is organized as follows. In section 2 we discuss an auxiliary
Plateau problem in d+1 dimensional Euclidean space. Already in two dimensions
the full analysis of the Plateau problem is quite nontrivial as it was
demonstrated in classical work of Douglas published in 1939 $[38]$.
Multidimensional treatment of this problem is even more nontrivial and touches
many subtle aspects of the harmonic analysis $[39].$ Nevertheless, the
extension of the Euclidean variant of the Plateau problem to the hyperbolic
H$^{d+1}$space is actually not difficult and was accomplished rather long time
ago by Ahlfors$[25]$. Using the results of Ahlfors we were able to reobtain
the results of Freedman et al, Ref.[12], almost straightforwardly in section
3. We deliberately considering only the scalar field case in this work since
the extension of our treatment to vector and tensor fields (to be briefly
considered in Section 8 ) does not cause much additional \textbf{conceptual}
problems. To generalize the results of section 3 and to put them in an
appropriate mathematical context, we discuss (in section 4) diffusion in the
hyperbolic space. This is done with several purposes. First, using symmetries
of the Laplace operator acting in the hyperbolic space it is possible to
subdivide Brownian motions on transient and recurrent. Only transient motions
can reach the boundary of the hyperbolic space. The transience and/or
recurrence is associated with convergence /or divergence of certain infinite
sums known as Poincar$e^{\prime}$ series.The convergence or divergence of such
series is being controlled by the critical exponent $\alpha.$ Patterson$[40]$,
Sullivan$[41]$, Ahlfors[25], Thurston $[42]$ and others $[32]$ had shown that
this exponent is associated with the fractal dimension of the limit set
$\Lambda.$ Stated in physical terms, it is shown in section 5 that this
exponent is associated with the exponent 2$\nu$ for the two-point correlation
function of the corresponding boundary CFT . The exponent $\alpha$ depends
upon the specific group of motions in H$^{d+1}$. This group is directly
associated with the hyperbolic manifold so that \textbf{different groups}
associated with \textbf{different} \textbf{manifolds} will produce
\textbf{different} $\alpha^{\prime}s.$ Being armed with these ideas it is
possible to improve the existing physical results using spectral theory of
hyperbolic manifolds in section 6. In this section it is shown that the
obtained eigenvalue spectrum of the hyperbolic Laplacian discussed in physics
literature is incomplete and much more results could be obtained with help of
the existing mathematical literature, e.g.see Ref.[28]. For instance, 2d
critical exponent 2$\nu$ for the Ising model is almost straightforwardly
obtained with help of the recently obtained results of Bishop and Jones$[43]$
.With this result obtained, it is only natural to look for connections between
the boundary CFT results and those coming from the fundamental work of
BPZ$[35]$ .The connection can be established rather easily,e.g.see section 7
,based on the theory of deformation of Kleinian groups$[27,31]$ which is
closely associated with the theory of Teichm\"{u}ller spaces$[44]$ as it was
demonstrated by Bers$[45]$ some time ago. One of the sources which generates
''new'' Kleinian groups from the ''old'' ones is through the extension of the
quasiconformal deformations produced at the boundary $\Omega=S_{\infty}%
^{2}-\Lambda$ of the hyperbolic space into the bulk(i.e.holography in physical
terminology). The theory of such deformations was under development in
mathematics for quite some time$.$ However, the results which are essential
for making connections with current physics literature had been obtained by
mathematicians only quite recently. In particular, Canary and Taylor$[46]$ had
demonstrated that the limit set of Kleinian groups which produce critical
exponents $\alpha$ in physically interesting range (e.g. for 0%
%TCIMACRO{\TEXTsymbol{<}}%
%BeginExpansion
$<$%
%EndExpansion
$\alpha<1$ one obtains the correct Ising model critical exponent
2$\nu=1/4,etc.)$ is a circle $S$, perhaps, with some points(or, may be,
segments) being removed (e.g. see section 7 for more details). These facts
naturally explain the crucial role being played by the loop groups and the
loop algebras$[34]$ in the conformal field theories and other exactly
integrable systems [47]. At the same time, Nag and Verjovsky$[48]$ had
demonstrated how the boundary deformations of such circle is connected with
the central extension term of the Virasoro algebra thus providing
\textbf{major physical reasons} for existence of such term. Moreover, the
analysis of the seminal work by Nag and Verjovsky indicates that, actually,
their main results are based entirely on much earlier work by Ahlfors[49] .
The Virasoro algebra and all results of the CFT [$35]$ could be obtained much
earlier should work by Ahlfors$[49],$ written in 1961, be properly interpreted
at that time. Ahlfors and many others (e.g.see Ref. [27] for a review) had
developed extension of theory of \ 2 dimensionalquasiconformal deformations to
hyperbolic spaces of higher dimensions. When these results are being put in a
proper physical context they allow extension of the BPZ formalism to higher
dimensions. The possibility of such extension(s) is discussed in section 8.
Taking into account that the conformal group in d dimensions is isomorphic to
the Lie group O(d+1,1) as noticed by Cartan in 1920's [50], for d=2 we obtain
the Lie group O(3,1) known also as Lorentz group. The connected part of this
group is isomorphic to $PSL(2,C)$ [51]. The Lie algebra of this group lies in
the center ($VectS^{1})$ of the Virasoro algebra . The central extension of
this algebra is just the Virasoro algebra. For d=3 \ we have the Lie group
O(4,1) known as de Sitter group. The representations of the Lie algebra for
this group ,fortunately, were studied both in mathematics [52,53] and in
physics[54,55] in connection with exact algebraic solution of the hydrogen
atom.Since the hydrogen atom is exactly solvable quantum mechanical problem,
construction of representations of the Lie algebra for the de Sitter Lie group
is also known. It is facilitated by the major observation[52,53] that the Lie
algebra of the de Sitter group can be presented as direct tensor product \ of
the Lie algebras for the group SO(3)$\simeq PSL(2,C).$ Hence, it is possible
to construct the central extensions for \textbf{each} of the Lie algebras
so(3) \textbf{independently }thus forming two copies of Virasoro algebras with
\textbf{different} central charges in general. Construction of the tensor
products of \ Virasoro algebras had been discussed in the literature already
(e.g. see Lecture 12 of Ref.[56]). This possibility is worth discussing only
if the limit set $\Lambda$ is \textbf{union} of two independent circles. Since
this fact had not been proven, to our knowlege, other possibilities also
exist, e.g. $\Lambda$ is still a circle. These possibilities are discussed
briefly in the same section. Recently, Bakalov, Kac and Voronov[57] were able
to extend the cohomological analysis of Gelfand and Fuks[58] \ thus obtaining
the higher dimensional analogue of the Virasoro conformal algebra (e.g. see
section 10 of Ref.[57]). It remains a challenging problem to recover these
results by developing the Kodaira-Spencer-like cohomological theory of
multidimensional quasiconformal deformations. Some efforts in this direction
are mentioned in the same section.

\medskip\mathstrut

{\Large 2. \ The Plateau problem in d+1 dimensional Euclidean space}

\mathstrut

The classical Plateau problem, when stated mathematically, essentially
coincides with the Dirichlet problem. In two dimensions the Dirichlet problem
can be formulated as follows:among functions $\varphi(z)$, z $\in A$ (where
$A$ is some closed domain of the complex plane $\mathbf{C)}$ which take values
$\varphi_{0}(z)$ at $\partial A$ find such that the Dirichlet integral
$D$[$\varphi]$ defined by
\begin{equation}
D[\varphi]=\iint\limits_{A}d^{2}z(\vec{\nabla}\varphi\cdot\vec{\nabla}\varphi)
\tag{2.1}%
\end{equation}
has the lowest possible value. Evidently, the above problem can be reduced to
the problem of finding the harmonic function $\varphi(z)$, i.e. the function
which obeys the Laplace equation
\begin{equation}
\Delta\varphi=0\text{ if z}\in A\text{ but z}\notin\bar{A} \tag{2.2}%
\end{equation}
and takes at the boundary $\partial A$ the preassigned values
\begin{equation}
\varphi\mid_{\partial A}=\varphi_{0}(z)\text{ .} \tag{2.3}%
\end{equation}
If $G(z,z^{\prime})$ is the Green's function of the Laplace operator $\Delta$,
then the harmonic function which possess the above properties is given by the
following boundary integral
\begin{equation}
\varphi(z)=-\int\limits_{\partial A}d\sigma\varphi_{0}(\sigma)\frac{\partial
G}{\partial n} \tag{2.4}%
\end{equation}
with normal derivative taken with respect to the direction of the exterior
normal.Use of Green's formulas allows one to rewrite the Dirichlet integral in
the following equivalent form
\begin{equation}
D[\varphi]=\iint\limits_{A}d^{2}z(\vec{\nabla}\varphi\cdot\vec{\nabla}%
\varphi)=\int\limits_{\partial A}d\sigma\varphi_{0}(\sigma)\frac
{\partial\varphi}{\partial n}\mid_{z=\sigma}. \tag{2.5}%
\end{equation}
By combining Eq.s (2.4) and (2.5) we obtain,
\begin{equation}
D[\varphi]=-\int\limits_{\partial A}d\sigma\varphi_{0}(\sigma)\int
\limits_{\partial A}d\sigma^{^{\prime}}\varphi_{0}(\sigma^{\prime}%
)\frac{\partial^{2}G}{\partial n\partial n^{\prime}}\text{ \quad.} \tag{2.6}%
\end{equation}
Taking into account that
\begin{equation}
\int\limits_{\partial A}d\sigma\frac{\partial G}{\partial n}=1 \tag{2.7}%
\end{equation}
which implies
\begin{equation}
\frac{\partial}{\partial n^{\prime}}\int\limits_{\partial A}d\sigma
\frac{\partial G}{\partial n}=0 \tag{2.8}%
\end{equation}
we can rewrite Eq.(2.6) in the following equivalent form
\begin{equation}
D[\varphi]=\frac{1}{2}\int d\sigma\int d\sigma^{\prime}[\varphi_{0}%
(\sigma)-\varphi_{0}(\sigma^{\prime})]^{2}\frac{\partial^{2}G}{\partial
n\partial n^{\prime}}\text{ \quad.} \tag{2.9}%
\end{equation}
Eq.(2.9) was derived by Douglas$[38]$ in 1939 in connection with his extensive
study of the Plateau problem and serves as starting point of all further
investigations related to two dimensional Plateau problem.

In the case if $\partial A$ is extended (long enough) contour, following
Douglas, we can use the Green's function for the half space given by
\begin{equation}
G(z,z^{\prime})=-\frac{1}{4\pi}\ln\frac{(x-x^{\prime})^{2}+(y-y^{\prime})^{2}%
}{(x-x^{\prime})^{2}+(y+y^{\prime})^{2}} \tag{2.10}%
\end{equation}
with z=x+iy, y%
%TCIMACRO{\TEXTsymbol{>}}%
%BeginExpansion
$>$%
%EndExpansion
0. To get $\frac{\partial^{2}G}{\partial n\partial n^{\prime}}$ we have to
keep only the infinitesimal values of y and y' in Eq.(2.10).This then
produces,
\begin{equation}
G(z,z^{\prime})\approx-\frac{1}{\pi}\frac{yy^{\prime}}{(x-x^{\prime})^{2}%
}\text{ ,} \tag{2.11}%
\end{equation}
so that
\begin{equation}
\frac{\partial^{2}G}{\partial n\partial n^{\prime}}=\frac{1}{\pi}\frac
{1}{(x-x^{\prime})^{2}}\text{ .} \tag{2.12}%
\end{equation}
Using this result in Eq.(2.9) we obtain,
\begin{equation}
D[\varphi]=\frac{1}{2\pi}\int d\sigma\int d\sigma^{\prime}[\varphi_{0}%
(\sigma)-\varphi_{0}(\sigma^{\prime})]^{2}\frac{1}{(\sigma-\sigma^{\prime
})^{2}}\text{ .} \tag{2.13}%
\end{equation}
This result is manifestly nonsingular for the well behaved function
$\varphi_{0}(\sigma)$ . The requirements on $\varphi_{0}(\sigma)$ needed for
$D$[$\varphi]$ to be nondivergent could be found in the already cited paper by
Douglas$[38].$In anticipation of physical applications, obtained results can
be easily extended now to higher dimensions. To do so, the metric of the
underlying space should be specified . Below we develop our results for the
case of Euclidean spaces of dimension d+1 while in the next section we shall
extend these results to the case of hyperbolic(Lobachevski) space H$^{d+1}$.
In the case of d+1 Euclidean space it is sufficient$[39]$ to consider the
Dirichlet problem for the half-space :\{x,z $\mid$ z%
%TCIMACRO{\TEXTsymbol{>}}%
%BeginExpansion
$>$%
%EndExpansion
0\}so that d$^{d+1}x=d^{d}xdz$ and $\varphi(x)=\varphi(\mathbf{x,}z)$ with
$\varphi_{0}(\mathbf{x})\equiv\varphi(\mathbf{x},0)$ or, equivalently, in the
unit d+1 dimensional ball B$^{d+1}$. An analogue of the Poisson formula,
Eq.(2.4), is known $[39]$ to be
\begin{equation}
\varphi(\mathbf{x},z)=\int\limits_{\partial A}d^{d}\mathbf{x}\text{ P}%
_{E}(z,\mathbf{x-x}^{\prime})\varphi_{0}(\mathbf{x}) \tag{2.14}%
\end{equation}
with
\begin{equation}
\text{P}_{E}(z,\mathbf{x-x}^{\prime})=c_{d+1}\frac{z}{[(\mathbf{x-x}^{\prime
})^{2}+z^{2}]^{\frac{d+1}{2}}} \tag{2.15}%
\end{equation}
where $c_{d+1}=\frac{2}{(d+1)V(B)}$ with
\begin{equation}
V(B)=\left\{
\begin{array}
[c]{c}%
\frac{\pi^{\frac{d+1}{2}}}{[(d+1)/2]!}\text{ if d+1 is even}\\
\frac{2^{\frac{d+2}{2}}\pi^{\frac{d}{2}}}{1\cdot3\cdot3\cdot\cdot\cdot
(d+1)}\text{ if d+1 is odd}%
\end{array}
\right.  \tag{2.16}%
\end{equation}
For example, if d+1=2 we obtain $c_{2}=\frac{1}{\pi}$ . This result is in
accord with Eq.(2.11) since using this equation and prescription of Douglas
$[38]$ we obtain,
\begin{equation}
\text{P}_{E}(z,\mathbf{x-x}^{\prime})=\frac{\partial}{\partial n}G=\frac
{1}{\pi}\frac{z}{z^{2}+(\mathbf{x-x}^{\prime})^{2}}\text{ .} \tag{2.17}%
\end{equation}
By repeating the same steps as in two dimensional case ,we obtain now the
following value for the Dirichlet integral
\begin{equation}
D[\varphi]=\frac{c_{d+1}}{2}\int\limits_{\partial A}d^{d}x\int
\limits_{\partial A}d^{d}x^{\prime}[\varphi_{0}(\mathbf{x})-\varphi
_{0}(\mathbf{x}^{\prime})]^{2}\frac{1}{\left|  \mathbf{x}-\mathbf{x}^{\prime
}\right|  ^{d+1}}\text{ .} \tag{2.18}%
\end{equation}
This result coincides with earlier obtained, Eq.(2.13), for the case of two
dimensions as required. Evidently, it could be made nonsingular if the
boundary function $\varphi_{0}(x)$ is appropriately chosen . Eq.(2.18) differs
from that known in physical literature, e.g. see Ref.[59] where, instead, the
following value for the Dirichlet integral was obtained
\begin{equation}
D[\varphi]=a_{d}\int\limits_{\partial A}d^{d}x\int\limits_{\partial A}%
d^{d}x^{\prime}\frac{\varphi_{0}(\mathbf{x})\varphi_{0}(\mathbf{x}^{\prime}%
)}{\left|  \mathbf{x}-\mathbf{x}^{\prime}\right|  ^{d+1}} \tag{2.19}%
\end{equation}
with constant $a_{d}$ left unspecified. Such integral could be potentially
divergent, unlike that given by Eq.(2.18), and, therefore, provides no
acceptable solution to the Dirichlet (or Plateau) problem in any dimension.
Obtained results can be easily generalized to the case of hyperbolic space.
This generalization is being treated in the next section.

\mathstrut

{\Large 3. \ The Plateau problem in d+1 dimensional hyperbolic space}

\mathstrut

Since the Euclidean variant of the AdS space is just a normal hyperbolic space
H$^{d+1}$ as was noticed in Ref.[13], we shall treat only the hyperbolic
Dirichlet (Plateau) problem in this paper. This is justified by the fact that
all results obtained in this work are in agreement with those obtained in
physics literature with help of less mathematically rigorous methods. Such an
agreement is not totally coincidental since it follows from deep results
obtained by Scannell[60] which provide a unified description of hyperbolic, de
Sitter and AdS spaces.

As it was shown by Ahlfors$[25]$ the Green's formulas of harmonic analysis
survive transfer to the hyperbolic space with minor modifications. For
example, for arbitrary (but well behaved) functions $u$ and $\ v$ the Green's
formula analogue for the hyperbolic space is given by
\begin{equation}
\int\limits_{V}u\Delta_{h}vd_{h}x=\int\limits_{\partial V}u\frac{\partial
v}{\partial n_{h}}\cdot d\sigma_{h}-\int\limits_{V}(\vec{\nabla}_{h}u\cdot
\vec{\nabla}_{h}v)dx_{h}\text{ .} \tag{3.1}%
\end{equation}
In particular, if $u=v$ and $u$ is hyperharmonic, i.e.
\begin{equation}
\Delta_{h}u=0\text{ \quad in }V\text{ } \tag{3.2}%
\end{equation}
then,
\begin{equation}
D[u]=\int\limits_{V}dx_{h}(\vec{\nabla}_{h}u\cdot\vec{\nabla}_{h}%
u)=\int\limits_{\partial V}u\frac{\partial u}{\partial n_{h}}\cdot d\sigma_{h}
\tag{3.3}%
\end{equation}
which is the hyperbolic analogue of Eq.(2.5). The subscript h in all above
equations stands for ''hyperbolic''. In particular, in case of B$^{d+1}$(d+1
dimensional ball of unit radius) model of hyperbolic space we have for the
hyperbolic Laplacian the following result
\begin{equation}
\Delta_{h}f(r)=\frac{1}{4}(1-r^{2})^{2}[\Delta f+\frac{2(d-1)}{1-r^{2}}%
r\frac{\partial f}{\partial r}] \tag{3.4}%
\end{equation}
with $r=\left|  x\right|  $ , ( $\left|  x\right|  =\sqrt{\sum\limits_{i=1}%
^{d+1}x_{i}^{2}}$ ) and
\begin{equation}
\Delta f(r)=\frac{d^{2}}{dr^{2}}f+\frac{d}{r}\frac{df}{dr} \tag{3.5}%
\end{equation}
while in the case of upper half space realization of the hyperbolic space we
have as well
\begin{equation}
\Delta_{h}f(\mathbf{x,}z)=z^{2}[\Delta f-(d-1)\frac{1}{z}\frac{\partial
f}{\partial z}]\text{ , \quad z%
%TCIMACRO{\TEXTsymbol{>}}%
%BeginExpansion
$>$%
%EndExpansion
0.} \tag{3.6}%
\end{equation}
It can be easily shown$[32],$ that for the upper half space model the
following eigenfunction equation holds
\begin{equation}
\Delta_{h}z^{\alpha}=\alpha(\alpha-d)z^{\alpha} \tag{3.7}%
\end{equation}
so that the function $z^{d}$ is hyperharmonic since it obeys the hyperharmonic
generalization of the Laplace Eq. (2.2):
\begin{equation}
\Delta_{h}z^{d}=0. \tag{3.8}%
\end{equation}
In the case of B$^{d+1}$ model we have as well $[25]$
\begin{equation}
dx_{h}=\frac{2^{d+1}dx_{1}dx_{2}...dx_{d+1}}{(1-\left|  x\right|  ^{2})^{d+1}%
}\text{ ,} \tag{3.9}%
\end{equation}%
\begin{equation}
d\sigma_{h}=\frac{2^{d}dx_{1}dx_{2}...dx_{d}}{(1-\left|  x\right|  ^{2})^{d}%
}\text{ ,} \tag{3.10}%
\end{equation}%
\begin{equation}
\frac{\partial u}{\partial n_{h}}=\frac{1-\left|  x\right|  ^{2}}{2}%
\frac{\partial u}{\partial n}\text{ ,} \tag{3.11}%
\end{equation}%
\begin{equation}
\vec{\nabla}_{h}u=\frac{1-\left|  x\right|  ^{2}}{2}\vec{\nabla}u\text{ .}
\tag{3.12}%
\end{equation}
The analogous formulas could be obtained for the H$^{d+1}$ model as well. The
hyperbolic Laplacian $\Delta_{h}$ possesses very important property of
M$\ddot{o}$bius invariance which can be formulated as follows. Let $\gamma
x=x^{\prime}$ be M$\ddot{o}$bius transformation of the hyperbolic space, i.e.
let $\gamma\in\Gamma$ where $\Gamma$ is the group of isometries which leave
H$^{d+1}$(or B$^{d+1})$ invariant then, for any function f, such that
\begin{equation}
\Delta_{h}f(x)=F(x) \tag{3.13a}%
\end{equation}
we have as well
\begin{equation}
\Delta_{h}f(\gamma x)=F(\gamma x). \tag{3.13b}%
\end{equation}
In particular, if the function $f(x)$ is hyperharmonic, then the function
$f(\gamma x)$ is also hyperharmonic. We have already mentioned, e.g. Eq.(3.7)
,that the function z$^{d}$ is hyperharmonic. Now we would like to use the
property of the hyperharmonic Laplacian given by Eq.(3.13b) in order to obtain
more general form of the hyperharmonic function in H$^{d+1}$. Using known
results for M$\ddot{o}$bius transformations in H$^{d+1}$ one easily obtains
(with accuracy up to unimportant constant)
\begin{equation}
f(x)=\left[  \frac{z}{\left|  z^{2}+(\mathbf{x}-\mathbf{x}^{\prime}%
)^{2}\right|  ^{2}}\right]  ^{d}\text{ .} \tag{3.14}%
\end{equation}
Let us check this result for the case of two dimensions first. In this case
d=1 in Eq.(3.14) and we obtain (with accuracy up to constant) Eq.(2.17). This
fact is not totally coincidental since, in view of Eq.(3.6), the hyperbolic
Laplacian coincides with the usual one for d=1. Therefore, we can write as
well in d+1 dimensions:
\begin{equation}
P_{H}(z,\mathbf{x}-\mathbf{x}^{\prime})=\hat{c}_{d}\left[  \frac{z}{\left|
z^{2}+(\mathbf{x}-\mathbf{x}^{\prime})^{2}\right|  ^{2}}\right]  ^{d},
\tag{3.15}%
\end{equation}
to be compared with Eq.(2.15). To calculate the constant $\hat{c}_{d}$ we have
to use known general properties of the Poisson kernels$[39].$ In particular,
the normalization requirement
\begin{equation}
\hat{c}_{d}\int d^{d}x\left[  \frac{z}{\left|  z^{2}+\mathbf{x}^{2}\right|
^{2}}\right]  ^{d}=1 \tag{3.16}%
\end{equation}
makes $P_{H}$ to act as probability density. This fact is going to be
exploited below.

Using spherical system of coordinates we easily obtain:
\begin{equation}
\hat{c}_{d}^{-1}=\omega_{d}\int\limits_{0}^{\infty}dx\frac{x^{d-1}}%
{(x^{2}+1)^{d}}=\frac{\omega^{d}}{2}\frac{\Gamma(\frac{d}{2})\Gamma(\frac
{d}{2})}{\Gamma(d)}, \tag{3.17}%
\end{equation}
where $\omega_{d}$ is the surface area of d-dimensional unit sphere ,
\begin{equation}
\omega_{d}=\frac{2\pi^{\frac{d}{2}}}{\Gamma(\frac{d}{2})}\text{ .} \tag{3.18}%
\end{equation}
By combining this result with Eq.(3.17) we obtain,
\begin{equation}
\hat{c}_{d}=\frac{\Gamma(d)}{\pi^{\frac{d}{2}}\Gamma(\frac{d}{2})}\text{ .}
\tag{3.19}%
\end{equation}
Given the results above, to obtain the Dirichlet integral using Eq.(3.3) is
rather straightforward, especially, by working in H$^{d+1}$ space. In this
case we have to replace Eq.s (3.9)-(3.12) by the following equivalent
expressions:
\begin{equation}
d\sigma_{h}=\frac{d^{d}x}{x_{0}^{d}}\text{ } \tag{3.20}%
\end{equation}
and
\begin{equation}
\frac{\partial u}{\partial n_{h}}=x_{0}\frac{\partial u}{\partial x_{0}}
\tag{3.21}%
\end{equation}
while keeping in mind Eq.(3.16). With these remarks we obtain at once
\begin{equation}
D[\varphi]=-d\hat{c}_{d}\int d^{d}x\int d^{d}x^{\prime}\frac{\varphi
_{0}(\mathbf{x})\varphi_{0}(\mathbf{x}^{\prime})}{\left|  \mathbf{x}%
-\mathbf{x}^{\prime}\right|  ^{2d}}\text{ .} \tag{3.22}%
\end{equation}
This result coincides with that obtained by Freedman et al in Ref[12]and,
later, in Ref.[15]. In both cases the methods which were used are noticeably
different fro ours.

From the discussion presented in section 2 it is clear that this result can be
rewritten in a manifestly nonsingular way thus removing need for
renormalization advocated in Ref.[14]. Actually, there is much more to it as
we shall demonstrate shortly below.

\mathstrut

{\Large 4. Diffusion in the hyperbolic space and boundary CFT }

{\Large \mathstrut}

The connection between the Klein-Gordon (K-G) and the Schr\"{o}dinger
propagators had been discussed already by Feynman long time ago and had been
exploited recently in our work, Ref.[61]. For reader's convenience,we would
like to repeat here these simple arguments. To this purpose,let us consider
the equation for K-G propagator in Euclidean space first. We have
\begin{equation}
\left(  \Delta-m^{2}\right)  G(\mathbf{x},\mathbf{x}^{\prime})=\delta
^{d}(\mathbf{x}-\mathbf{x}^{\prime})\text{ .} \tag{4.1}%
\end{equation}
By introducing the fictitious (or real) time variable t the auxiliary
equation
\begin{equation}
\frac{\partial}{\partial t}\hat{G}(\mathbf{x},\mathbf{x}^{\prime};t)=\left(
\Delta-m^{2}\right)  \hat{G}(\mathbf{x},\mathbf{x}^{\prime};t) \tag{4.2}%
\end{equation}
supplemented with the initial condition
\begin{equation}
\hat{G}(\mathbf{x},\mathbf{x}^{\prime};t=0)=\delta^{d}(\mathbf{x}%
-\mathbf{x}^{\prime}) \tag{4.3}%
\end{equation}
is useful to consider in connection with Eq.(4.1). The correctness of the
initial condition could be easily cheked. Indeed, since the solution of
Eq.(4.2) is known to be
\begin{equation}
\hat{G}(\mathbf{x},\mathbf{x}^{\prime};t)=\int\frac{d^{d}k}{\left(
2\pi\right)  ^{d}}\exp\{-i\mathbf{k}\cdot(\mathbf{x}-\mathbf{x}^{\prime
})-t(\mathbf{k}^{2}+m^{2})\} \tag{4.4}%
\end{equation}
one obtains immediately the result given by Eq.(4.3). At the same time, if the
solution of Eq.(4.2) is known then, the solution of Eq.(4.1) is known as well
and is given simply by
\begin{equation}
G(\mathbf{x},\mathbf{x}^{\prime})=\int\limits_{0}^{\infty}dt\hat{G}%
(\mathbf{x},\mathbf{x}^{\prime};t)\text{ .} \tag{4.5}%
\end{equation}
One can do even better by noticing that the mass term in Eq.(4.2) can be
simply eliminated by using the following substitution:
\begin{equation}
\hat{G}(\mathbf{x},\mathbf{x}^{\prime};t)=e^{-m^{2}t}\tilde{G}(\mathbf{x}%
,\mathbf{x}^{\prime};t)\text{ .} \tag{4.6}%
\end{equation}
Thus introduced function $\tilde{G}$ obeys the standard diffusion equation:
\begin{equation}
\frac{\partial}{\partial t}\tilde{G}(\mathbf{x},\mathbf{x}^{\prime}%
;t)=\Delta\tilde{G}(\mathbf{x},\mathbf{x}^{\prime};t)\text{ .} \tag{4.7}%
\end{equation}
which is just the Euclidean version of The Schrodinger equation for the free
particle propagator. From the theory of random walks it is well known$[62]$
that in the case of $m^{2}=0$ and $\mathbf{x}=\mathbf{x}^{\prime}$ the
quantity
\begin{equation}
G(\mathbf{0})=\int\limits_{0}^{\infty}dt\hat{G}(\mathbf{0};t) \tag{4.8}%
\end{equation}
represents the average time $\ <T>$ \quad which Brownian particle spends at
the origin(initial point).The probability $\Pi(\mathbf{0})$ of returning to
the origin is known to be related to $G(0)$ as follows$[62]$%
\begin{equation}
\Pi(\mathbf{0})=1-\frac{1}{G(\mathbf{0})}\text{ .} \tag{4.9}%
\end{equation}
Accordingly,the random walk is \textbf{recurrent }or \textbf{transient}
depending upon $\Pi(\mathbf{0})$ being equal to or lesser than one. The
''recurrent'' means that the ''particle'' will come to the origin time and
again while the ''transient'' means that \textbf{finite} probability it will
leave the origin and may never come back.

Thus, from the point of view of the theory of Brownian motion, the Dirichlet
problem discussed in sections 2 and 3 is associated with the question about
the probability for the random walker to reach the boundary $S_{\infty}^{d}$
(in the case of B$^{d+1}\operatorname{mod}$el)or $R^{d}$ (in the case of
H$^{d+1}$model)of the hyperbolic space or, alternatively, the random walk must
be \textbf{transient} in order to be able to reach the boundary. This can be
formulated also as the condition
\begin{equation}
G(\mathbf{0})<\infty\tag{4.10}%
\end{equation}
for the Dirichlet problem to be well posed. This condition may or may not be
fulfilled as we shall discuss shortly. In the meantime, we would like to
return to the massive case in order to extend to this case the above described
concepts. Using Eq.(3.7) we obtain now for the massive case the following
requirement
\begin{equation}
\alpha(\alpha-d)-m^{2}=0 \tag{4.11}%
\end{equation}
for the function $z^{\alpha}$ to remain hyperharmonic. Eq.(4.11) leads to the
following values of $\alpha$ :
\begin{equation}
\alpha_{1,2}=\frac{d}{2}\pm\sqrt{\left(  \frac{d}{2}\right)  ^{2}+m^{2}}\text{
.} \tag{4.12}%
\end{equation}
To determine which of the values of $\alpha$ are acceptable, it is sufficient
only to check the normalization condition analogous to that used in Eq.(3.16).
To this purpose the Poisson-like formula (e.g. see Eq.(2.14)) is helpful. In
the present case we have
\begin{equation}
\varphi(\mathbf{x},z)=\hat{c}_{\alpha}\int\limits_{R^{d}}d^{d}x\left[
\frac{z}{\left(  x-x^{\prime}\right)  ^{2}+z^{2}}\right]  ^{\alpha}\varphi
_{0}(x^{\prime})\text{ .} \tag{4.13}%
\end{equation}
If $\alpha=d$ , then it is easy to see that for $\varphi_{0}(\mathbf{x}%
)$=const the r.h.s. of Eq.(4.13) is z-independent. If $\alpha\neq d$ , then
after rescaling: x$\rightarrow\frac{x}{z}\equiv y$ ,we are left with the
factor $z^{d-\alpha}$ under the integral.This factor can be eliminated if we
require
\begin{equation}
z^{d-\alpha}\varphi_{0}(yz)=\varphi_{0}(y). \tag{4.14}%
\end{equation}
This provides the boundary field $\varphi_{0}$ with the scaling dimension
$\Delta_{0}=d-\alpha$ in complete accord with Ref.[12] where this result was
obtained by use of slightly different set of arguments.

Now we are in the position to determine the actual value of the constant
$\hat{c}_{\alpha}.$ By analogy with Eq.(3.17) we obtain,
\[
\hat{c}_{\alpha}^{-1}=\omega_{d}\int\limits_{0}^{\infty}dx\frac{x^{d-1}%
}{\left(  x^{2}+1\right)  ^{\alpha}}=\frac{\omega_{d}}{2}\frac{\Gamma
(\alpha-\frac{d}{2})\Gamma(\frac{d}{2})}{\Gamma(\alpha)}%
\]
or, alternatively,
\begin{equation}
\hat{c}_{\alpha}=\frac{\Gamma(\alpha)}{\pi^{\frac{d}{2}}\Gamma(\alpha-\frac
{d}{2})}\text{ .} \tag{4.15}%
\end{equation}
For $\alpha=\frac{d}{2}$ the above equation becomes singular. This observation
leaves us with an option of choosing ''+'' sign in Eq.(4.12). This option, is
not the only one as it will be demonstrated below. In addition, the mass
$m^{2}$ should be larger than -$\left(  \frac{d}{2}\right)  ^{2}$ for the sake
of the normalization requirement. These conclusions coincide with the results
of Sullivan $[41]$ who reached them by using a somewhat different set of
arguments. Using Eq.(4.13) and repeating the same steps as in the massless
case, e.g.see Eqs.(3.20)-(3.22), we obtain for the Dirichlet integral the
following final result:
\begin{equation}
D[\varphi]=-\hat{c}_{\alpha_{+}}\int d^{d}x\int d^{d}x^{\prime}\frac
{\varphi_{0}(\mathbf{x})\varphi_{0}(\mathbf{x}^{\prime})}{\left|
\mathbf{x}-\mathbf{x}^{\prime}\right|  ^{2\alpha_{+}}}\text{ .} \tag{4.16}%
\end{equation}
Eq.(4.16) is in formal agreement with results obtained in Refs[12],[15].
Unlike Refs[12],[15], where no further analysis of these results was made, we
would like to examine obtained results in more detail. As we had mentioned
already in the Introduction, according to Maskit$[8],$ the group of M$\ddot
{o}$bius transformations acts as a group of isometries in the hyperbolic space
H$^{d+1}($ or $B^{d+1})$ but \textbf{not} at its boundary. At the boundary of
the hyperbolic space it acts only as a group of conformal
''motions''(transformations) which is ''\textbf{not} a group of isometries in
any metric''$[8].$ \ If we take into account that the isometric motions in the
hyperbolic space are described by a group $\Gamma$ of M$\ddot{o}$bius
transformations, then Eq.s(4.5),(4.6) should be modified. In particular, we
should write, instead of Eq.(4.5), the following result:
\begin{equation}
G(\mathbf{x},\mathbf{x}^{\prime})=\sum\limits_{\gamma\in\Gamma}\int
\limits_{0}^{\infty}dte^{-m^{2}t}\tilde{G}(\mathbf{x},\gamma\mathbf{x}%
^{\prime};t). \tag{4.17}%
\end{equation}
The integral in Eq. (4.17) can be estimated ,e.g.see the discussion presented
in sections 5 and 6, and is roughly given by
\begin{equation}
\int\limits_{0}^{\infty}dte^{-m^{2}t}\tilde{G}(\mathbf{x},\gamma
\mathbf{x}^{\prime};t).\lesssim c\exp\{-\alpha_{+}\rho(\mathbf{x}%
,\gamma\mathbf{x}^{\prime})\} \tag{4.18}%
\end{equation}
where $\rho(\mathbf{x},\mathbf{x}^{\prime})$ is the hyperbolic distance
between $\mathbf{x}$ and $\mathbf{x}^{\prime}.$ It can be shown $[32],[41],$
that the convergence or divergence of the Poincar$e^{\prime}$ series
\begin{equation}
g_{\alpha_{+}}(\mathbf{x},\mathbf{x}^{\prime})=\sum\limits_{\gamma\in\Gamma
}\exp\{-\alpha_{+}\rho(\mathbf{x},\gamma\mathbf{x}^{\prime})\} \tag{4.19}%
\end{equation}
is actually independent of $\mathbf{x}$ and $\mathbf{x}^{\prime}$ . Hence, one
can choose as well both $\mathbf{x}$ and $\mathbf{x}^{\prime}$ at the center
of the hyperbolic ball $B^{d+1}$. Then, if the Poincare$^{\prime}$ series is
\textbf{divergent}, we have the recurrence (or ergodicity$[31],[41])$
according to Eq.(4.9), and if it is convergent, we have the transience. In
this case the random walk which had originated somewhere inside of the
hyperbolic space is going to end up its ''motion'' at the boundary of this
space. The exponent $\alpha_{+}$ responsible for this process of convergence
or divergence is associated with particular Kleinian (M$\ddot{o}$bius) group
$\Gamma$ so that different groups may have different exponents. To facilitate
reader's understanding, we would like to provide an introduction into these
very interesting topics in the next section.

\mathstrut

{\Large 5. The limit sets of Kleinian groups}\quad

\mathstrut

By definition, Kleinian groups are groups of isometries of H$^{3}$ (or
$B^{3})$ ,e.g.see Ref[8], while M$\ddot{o}$bius groups are groups of
isometries of H$^{d+1}$(or B$^{d+1})$ for d$\geq1.$ Hence, Kleinian groups are
just a special case of the M$\ddot{o}$bius groups. Recall also that Kleinian
groups are just complex version of Fuchsian groups acting on H$^{2}.$

Let $\Gamma$ be one of such groups and let $\gamma\in\Gamma$ be some
representative element of such group. For an arbitrary $x\in$H$^{d+1}$ the
group $\Gamma$ acts \textbf{discontinuously} if $\gamma x\cap x$ is nonempty
only for finitely many $\gamma\in\Gamma$ . In particular, the finite subgroup
$G_{0}$ is called \textbf{stabilizer }of the group $\Gamma$ if $gx^{\ast
}=x^{\ast}$ for $g\in G_{0}\in\Gamma$ and $x^{\ast}\in$H$^{d+1}$. The fixed
point(s) $x^{\ast}$ could be either inside of H$^{d+1}$ or at its boundary
$R^{d}$ . Every discontinuous group is also \textbf{discrete} $[63].$ A group
$\Gamma$ is \textbf{discrete }if there is no sequence $\gamma_{n}\rightarrow
I,n=1,2,...$ with \textbf{all} $\gamma_{n}$ being distinct. Discretness
implies that for any $x\in B^{d+1}$the orbit: $\gamma x,\gamma^{2}x,\gamma
^{3}x$ accumulates only at $S_{\infty}^{d}$ ,e.g.see Refs[10],[63],[64].

An orbit which has precisely one fixed point on $S_{\infty}^{d}$ is being
associated with the \textbf{parabolic} subgroup elements of $\Gamma$ while an
orbit which has two fixed points on $S_{\infty}^{d}$ is being associated with
the hyperbolic subgroup elements of $\Gamma$. Some important physical
applications of these definitions associated with Thurston's theory of
measured foliations and laminations had been recently discussed in
Refs[36],[37] in connection with description of dynamics of 2+1 gravity and
disclinations in liquid crystals.

There are also \textbf{elliptic} transformations but their fixed points always
lie \textbf{inside} of B$^{d+1}$and, therefore, are not of immediate physical
interest. The parabolic transformations are conjugate to translations
T:$x\rightarrow x+1($in the H$^{d+1}$ model these motions are motions in
R$^{d}$ which leave the ''time'' axis z unchanged ).The hyperbolic
transformations are conjugate to dilatations D:$x\rightarrow kx$ with $k>0$
and $k\neq1,$while the elliptic transformations are conjugate to rotations R:
$x\rightarrow e^{i\theta}x$ about the origin .

The question arises: how to describe the limit set $\Lambda$ of fixed points
which belong to $S_{\infty}^{d}?$ First, it is clear that, by construction,
$\Lambda$ is \textbf{closed} set since for all $x\in B^{d+1}$the orbit
\{$\gamma x\}\in\Lambda$ . Second, it can be shown $[64]$ that $\Lambda$ may
either contain no more than two points (\textbf{elementary} set$)$ or
uncountable number of points (\textbf{non- elementary} set ). In the last case
either $\Lambda=S_{\infty}^{d}$ or $\Lambda$ is nowhere dense in $S_{\infty
}^{d}.$ M\"{o}bius (or Kleinian) groups for which $\Lambda=S_{\infty}^{d}$ are
known as M\"{o}bius (or Kleinian) groups of the \textbf{first} kind while
M\"{o}bius (or Kleinian) groups for which $\Lambda\neq S_{\infty}^{d}$ are
known as groups of the \textbf{second} kind. The main goal of the subsequent
discussion is to provide enough evidence to the fact that the Green's function
for the hyperbolic Laplacian, Eq.(3.6), \textbf{exist if and only if }the
M$\ddot{o}$bius group $\Gamma$ is of \textbf{convergence type} (that is the
Poincar$e^{\prime}$ series, e.g. see Eq.(5.7) below, is convergent). In
Ref.[25] it is demonstrated that every Mobius group of the \textbf{second}
kind is of convergence type. This implies that the correlation function
exponent, e.g. see Eq.(4.16), is associated with the Hausdorff dimension of
the limit set $\Lambda$ which thus forms a fractal.

Let us begin with the fundamental property of the hyperbolic Laplacian
expressed in Eq.s (3.13a) and (3.13b). This property implies that in the
hyperbolic space $B^{d+1}$the Dirichlet (or Plateau) problem can be considered
\textbf{only} in conjunction with the group of motions (isometries) in this
space. In particular, let us consider an analogue of the Poisson formula,
Eq.(2.14), for the hyperbolic $B^{d+1}$ model. We have,
\begin{equation}
\varphi(x)=\frac{1}{\omega_{d}}\int\limits_{S_{\infty}^{d}}d\omega(x^{\prime
})\left(  \frac{1-\left|  x\right|  ^{2}}{\left|  x-x^{\prime}\right|  ^{2}%
}\right)  ^{d}\varphi_{0}(x^{\prime}) \tag{5.2}%
\end{equation}
where $d\omega$ is the $areal$ measure of $S_{\infty}^{d}$. Consider now a
special case of Eq.(5.2) when $\varphi_{0}(x)=const.$ Then, evidently,
$\varphi(x)=const$ too since the r.h.s. is constant by requirement of
normalization as it was discussed in section 3. This means, in turn, that
Eq.(4.16) does not exist for $\varphi_{0}(x)=const.$ Assume now that
$\varphi_{0}(x)$ is given by $\chi(x)$ with $\chi(x)$ being the
characterristic function of the set $\Lambda\in S_{\infty}^{d}.$ Let us assume
furthermore, in accord with definitioins provided earlier, that $\chi(\gamma
x)=\chi(x)$ (since the set $\Lambda$ is closed) with $\gamma\in\Gamma$ .Then,
using Eq.(5.2), we obtain,
\begin{equation}
\varphi(\gamma x)=\frac{1}{\omega_{d}}\int\limits_{S_{\infty}^{d}}%
d\omega(x^{\prime})\left(  \frac{1-\left|  \gamma x\right|  ^{2}}{\left|
\gamma x-\gamma x^{\prime}\right|  ^{2}}\right)  ^{d}\left|  \gamma^{\prime
}(x^{\prime})\right|  ^{d}\chi(x^{\prime})\text{ .} \tag{5.3}%
\end{equation}
But, since it is known $[25]$ that
\[
1-\left|  \gamma x\right|  ^{2}=\left|  \gamma^{\prime}(x)\right|  (1-\left|
x\right|  ^{2})
\]
and
\[
\left|  \gamma x-\gamma x^{\prime}\right|  ^{2}=\left|  \gamma^{\prime
}(x)\right|  \left|  \gamma^{\prime}(x^{\prime})\right|  \left|  x-x^{\prime
}\right|  ^{2}%
\]
where $\gamma^{\prime}(x)=\frac{d\gamma}{dx},$ we obtain,
\begin{equation}
\varphi(\gamma x)=\frac{1}{\omega_{d}}\int\limits_{S_{\infty}^{d}}%
d\omega(x^{\prime})\left(  \frac{1-\left|  x\right|  ^{2}}{\left|
x-x^{\prime}\right|  ^{2}}\right)  ^{d}\chi(x^{\prime})\text{ .} \tag{5.4}%
\end{equation}
That is
\begin{equation}
\varphi(\gamma x)=\varphi(x). \tag{5.5}%
\end{equation}
This means, that the function $\varphi(x)$ is authomorphic. Since the Poisson
kernel, in Eq.(5.4) is related to the corresponding Poisson kernel, Eq.(3.14),
in H$^{d+1}$ model, and, therefore, is related to the eigenfunction $z^{d}$ of
the hyperbolic Laplacian defined by Eq.s (3.7),(3.8), we conclude, that
$\varphi(x)$ is hyperharmonic and is nonconstant. This however, cannot be the
case for any nonzero areal measure, i.e. $\forall\chi(x)d\omega\neq0$ . To
understand why this is so several facts from the theory of fractals are
helpful at this point. Following Mandelbot$[65],$ let us recall the Olbers
paradox. Consider an observer in flat Euclidean Universe (which is assumed to
be 3 dimensional) located at some fixed point chosen as an origin. The amount
of light reaching an observer coming from some star located at distance$\sim
R$ is known to scale as $R^{-2}$. At the same time, if the density of stars is
roughly uniform, then the total mass of stars in the spherical volume of
radius $R$ is $\sim R^{3}$ so that the number of stars located at the visual
sphere of radius $R$ is $\sim R^{2}$ and, therefore, the amount of light
coming to observer is of order $\sim R^{2}\cdot R^{-2}=const.$ That is the sky
in such Euclidean Universe is uniformly lit day and night. This is, of course,
not true. The resolution of this paradox can be reached if one assumes that
the distribution of stars is that characteristic for fractals with the total
mass of stars on the visual sphere being $\sim R^{D}$ where the fractal
dimension $D<2.$ That this is indeed the case was demonstrated by
Sullivan$[66]$ (and, independently, by Tukia$[67])$ based on earlier work by
Thurston$[42]$ provided, that our Universe is \textbf{not} Euclidean but
Hyperbolic. Both Thurston and Sullivan were \textbf{not} concerned with Olbers
paradox but rather with the fractal dimension of the limit set $\Lambda$ which
is located at the sphere at infinity $S_{\infty}^{2}$ in $B^{2+1}$model of the
hyperbolic space. Using intuitive terminology, their results could be stated
as follows.

Let $B_{0}$ be some small ball located inside the hyperbolic space $B^{3}$ at
some point $a\in B^{3}$ . Let the noneuclidean radius $\rho$ of $B_{0}$be so
small that the images of $B_{0}$ given by $\gamma B_{0},\gamma^{2}%
B_{0},....,\gamma\in\Gamma,$ do not overlap. Instead of balls consider now
their ''shadows'' on $S_{\infty}^{2}$(as if inside of $B_{0}$ there is a
source of light which illuminates $B^{3}$ Universe). Denote $\gamma
B_{0}=B_{1},...,\gamma^{n}B_{0}=B_{n}$, etc., and, accordingly,for shadows,
$B_{1}^{^{\prime}},B_{2}^{^{\prime}},...,B_{n}^{^{\prime}}$ ,... \ Let now
L=$\bigcup\limits_{i}B_{i}^{^{\prime}}$ so that the areal measure
$\omega\equiv m(L).$

The Thurston- Ahlfors \textbf{Theorem} $[25]$ can now be informally stated as follows.

If \quad\quad\quad\quad\quad\quad\quad\quad\quad\quad$\sum\limits_{i=0}%
^{\infty}m(B_{i}^{^{\prime}})<\infty,$ \quad\quad\quad\quad\quad\quad
\quad\quad\quad\quad\quad\quad\quad\quad\quad\quad\quad

then, \quad\quad\quad\quad\quad\quad\quad\quad\quad$m(L)=0$ \quad and vice
versa. \quad\quad\quad\quad\quad\quad\quad\quad\quad\quad\quad\quad\quad\quad

The above is possible only if some of the shadows of the balls $B_{i}$ lie
completely (or partially) \textbf{inside }the shadows of other balls (located
closer to $B_{0}).$ The hard part of the proof of this theorem lies precisely
in proving that this \textbf{is} the case. We are not going to reproduce the
details of the proof in this paper(the reader is urged to consult
Refs[25],[31]section 9.9, for elegant and detailed proofs). Rather, we would
like to state the same results in more precise terms. This can be done by
noticing that, if
\begin{equation}
\int d\omega(x)\chi(x)=0, \tag{5.6}%
\end{equation}
then the Poincar$e^{\prime}$ series (e.g.see Eq.(4.19)) converges, that is
\quad%
\begin{equation}
\sum\limits_{\gamma\in\Gamma}\exp\{-\alpha\rho(\mathbf{x},\gamma
\mathbf{x}^{\prime})\}<\infty\tag{5.7}%
\end{equation}
and vice versa. Or, equivalently, if
\begin{equation}
\int d\omega(x)\chi(x)=\omega_{d}^{{}} \tag{5.8}%
\end{equation}
with $\omega_{d}$ being given by Eq.(3.18),then
\begin{equation}
\sum\limits_{\gamma\in\Gamma}\exp\{-\alpha\rho(\mathbf{x},\gamma
\mathbf{x}^{\prime})\}=\infty\text{ .} \tag{5.9}%
\end{equation}
Let us explain the obtained results in more physically familiar terms. First,
in view of the results of section 4, it is clear, that the results obtained
above could be equivalently stated in terms of recurrence(transience) of
random walks. Next, let us examine closer the Poisson kernel in Eq.(5.2), that
is
\begin{equation}
P_{H}^{\alpha}(x,x^{\prime})=\left(  \frac{1-\left|  x\right|  ^{2}}{\left|
x-x^{\prime}\right|  ^{2}}\right)  ^{\alpha} \tag{5.10}%
\end{equation}
where we had replaced $d$ in Eq.(5.2) by $\alpha$ for reasons which will
become clear shortly below. Notice, that $x^{\prime}\in S_{\infty}^{d}$ while
$x^{{}}\in B^{d+1}$ in Eq.(5.10). Consider the horoball centered at
$x^{\prime}\in S_{\infty}^{d}$ and passing through point $x\in B^{d+1}$as
depicted in Fig.1%

%TCIMACRO{\FRAME{ftbpFU}{2.3298in}{2.7942in}{0pt}{\Qcb{Some geometric relations
%in the hyperbolic ball model}}{}{FIG1.EPS}%
%width 2.3298in;  height 2.7942in;  depth 0pt;  original-width 2.6775in;
%original-height 4.2921in;  cropleft "0";  croptop "1";  cropright "1";
%cropbottom "0.2502";  filename 'FIG1.EPS';file-properties "XNPEU";}}}%
%BeginExpansion
\begin{figure}
[ptb]
\begin{center}
\includegraphics[
trim=0.000000in 1.073883in 0.000000in 0.000000in,
natheight=4.292100in,
natwidth=2.677500in,
height=2.7942in,
width=2.3298in
]%
{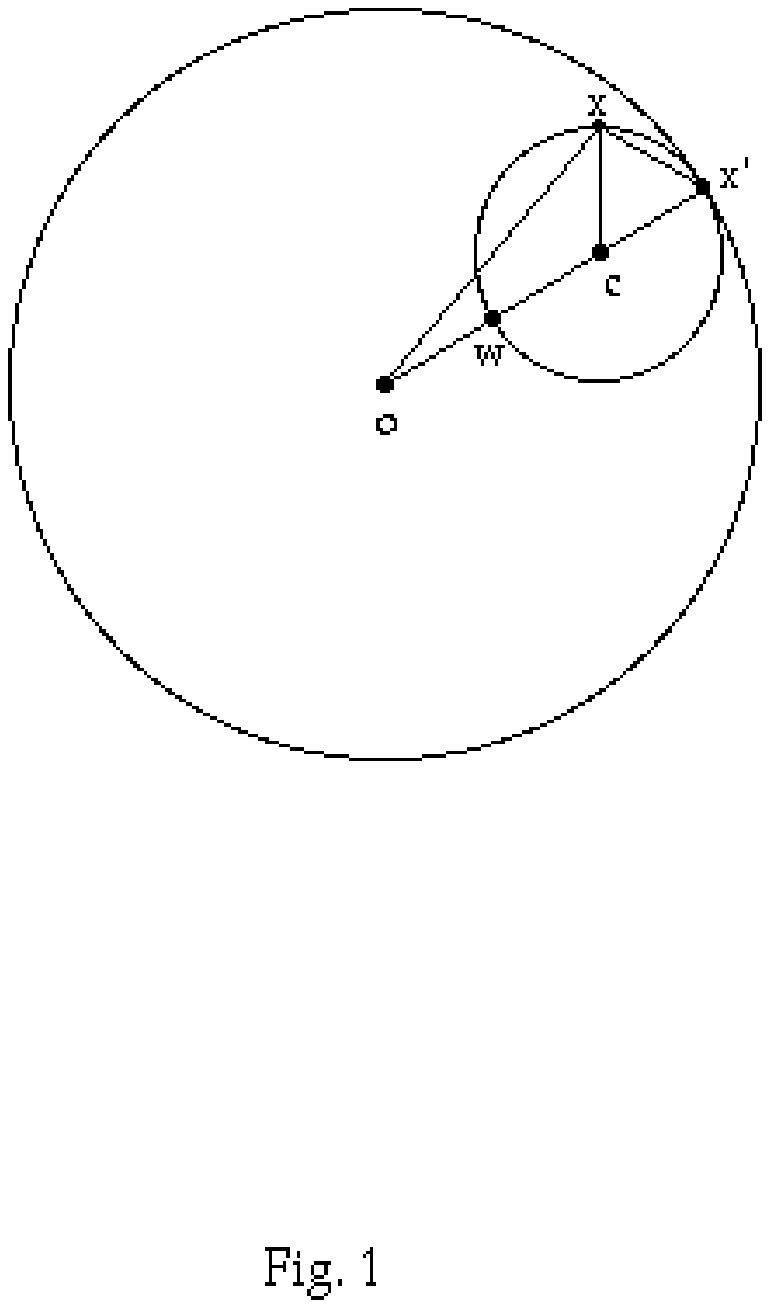}%
\caption{Some geometric relations in the hyperbolic ball model}%
\end{center}
\end{figure}
%EndExpansion

Using the cosine theorem for the angle $xox^{\prime}$ in the triangle
$\Delta_{xox^{\prime}}$ we obtain,
\begin{equation}
\left|  x\right|  ^{2}+1-2\left|  x\right|  \cos(xox^{\prime})=\left|
x-x^{\prime}\right|  ^{2}\text{ .} \tag{5.11}%
\end{equation}
\quad Alternatively, by using the triangle $\Delta_{xoc}$ we get
\begin{equation}
\left|  x\right|  ^{2}+\left|  w+\frac{1}{2}(1-w)\right|  ^{2}-2\left|
x\right|  \left|  w+\frac{1}{2}(1-w)\right|  \cos(xox^{\prime})=\frac{1}%
{4}(1-\left|  w\right|  )^{2}. \tag{5.12}%
\end{equation}
By eliminating $\cos(xox^{\prime})$ from these two equations we obtain,
\begin{equation}
\frac{1+\left|  x\right|  ^{2}-\left|  x-x^{\prime}\right|  ^{2}}{2}%
=1+\frac{\left|  x\right|  ^{2}-1}{1+\left|  w\right|  }\text{ .} \tag{5.13}%
\end{equation}
This result can be equivalently rewritten as
\begin{equation}
\frac{1-\left|  w\right|  }{1+\left|  w\right|  }=\frac{\left|  x-x^{\prime
}\right|  ^{2}}{\left|  x\right|  ^{2}-1}\text{ .} \tag{5.14}%
\end{equation}
The \textbf{hyperbolic} distance $\rho(0,w)$ is known to be $[63]$%
\begin{equation}
\rho(0,w)=\ln\left(  \frac{1+\left|  w\right|  }{1-\left|  w\right|  }\right)
\text{ .} \tag{5.15}%
\end{equation}
Accordingly, the Poisson kernel, Eq.(5.10), can be equivalently rewritten as
\begin{equation}
P_{H}(x,x^{\prime})=\exp\{\alpha\rho(0,w)\}. \tag{5.16}%
\end{equation}
The hyperbolic Fourier transform can be defined now as $[68]$
\begin{equation}
\varphi_{\alpha}(x)=\frac{1}{\omega_{d}}\int\limits_{S_{\infty}^{d}}%
d\omega(x^{\prime})\exp\{\alpha<x,x^{\prime}>\}\hat{\varphi}(x^{\prime})
\tag{5.17}%
\end{equation}
with scalar product $<x,x^{\prime}>$ being defined through the hyperbolic
distance $\rho(0,w)$ according to Eq.s (5.14),(5.15).

With help of the results just obtained it is possible to give better
interpretation of the Ahlfors-Thurston Theorem. Indeed, in view of
Eq.s(5.3)-(5.5) we obtain,
\begin{equation}
\varphi(0)=\frac{1}{\omega_{d}}\sum\limits_{\gamma\in\Gamma}\int
\limits_{S_{\infty}^{d}}d\omega(x^{\prime})\left(  \frac{1-\left|
\gamma(0)\right|  ^{2}}{\left|  \gamma(0)-x^{\prime}\right|  ^{2}}\right)
^{d}\chi(x^{\prime}), \tag{5.18}%
\end{equation}
where ,without loss of generality, we had put x=0 (i.e.placed the initial
point x at the center of $B^{d+1})$ .Surely, $\left|  \gamma(0)-x\right|
^{2}\leq4$ since we are dealing with the ball of unit radius. Therefore, we
also have
\begin{equation}
\varphi(0)<\frac{1}{\omega_{d}}\sum\limits_{\gamma\in\Gamma}\int
\limits_{S_{\infty}^{d}}d\omega(x^{\prime})(1-\left|  \gamma(0)\right|
^{2})^{d}\chi(x^{\prime})\text{ .} \tag{5.19}%
\end{equation}
Consider now the convergence(or divergence) of the series
\begin{equation}
S_{d}=\sum\limits_{\gamma\in\Gamma}(1-\left|  \gamma(0)\right|  ^{2})^{d}
\tag{5.20}%
\end{equation}
or, more generally,
\begin{equation}
S_{\alpha}=\sum\limits_{\gamma\in\Gamma}(1-\left|  \gamma(0)\right|
^{2})^{\alpha}\text{ .} \tag{5.21}%
\end{equation}
Clearly, the last expression is going to be divergent or convergent along
with
\begin{equation}
g_{\alpha}(0,0)=\sum\limits_{\gamma\in\Gamma}\left(  \frac{1-\left|
\gamma(0)\right|  }{1+\left|  \gamma(0)\right|  }\right)  ^{\alpha}%
=\sum\limits_{\gamma\in\Gamma}\exp\{-\alpha\rho(0,\gamma(0))\} \tag{5.22}%
\end{equation}
in view of Eq.s (5.15) and (4.19). But the convergence(divergence) of the
Poincar$e^{\prime}$ series , Eq.(5.22), leads us to the results given by
Eq.s(5.6)-(5.9). and also earlier stated result, Eq.(4.19).

The results just obtained admit yet another interpretation. Convergence (or
divergence) of the series, Eq.(5.22), is associated with existence or
nonexistence of the Green's function acting in $B^{d+1}$as we had mentioned
already beforeEq.(5.2)$.$ Deep results of Ahlfors$[25]$, Thurston$[42]$,
Patterson$[40]$, Beardon$[69]$ and Sullivan $[41]$ state that if the
Poincar$e^{\prime}$ series converges, then the Green's function in $B^{d+1}$
exist and the limit set $\Lambda\subset S_{\infty}^{d}$ is fractal with
$areal$ measure equal to zero but $Hausdorff$ dimension equal to $\alpha$ (in
this case $\alpha$ lies at the border between the convegence and divergence of
the series,Eq.(5.22))and, for d=2, $\alpha\leq2$ according to Sullivan$[66]$
and Tukia$[67]$ . Additional very important results related to the limit set
$\Lambda$ were obtained by Beardon and Maskit$[10]$ who had proved the following

\textbf{Theorem 5.1.(}Beardon Maskit\textbf{)} \textit{Let }$\Gamma
$\textit{\ be a discrete M}$\ddot{o}$\textit{bius group of isometries of
H}$^{3}$\textit{, then, if }$\Gamma$\textit{\ is \textbf{geometrically
finite}, the limit set }$\Lambda$\textit{\ comprizes of parabolic limit points
and conical limit points .}

We would like now to explain the physical significance and the meaning of
these statements. First, by looking at Eq.s(4.12)and (4.16) we conclude that
$m^{2}\leq0$ $($because of the results of Sullivan and Tukia). Second, for the
group $\Gamma$ to be \textbf{geometrically finite} ($in$ $B^{3})$ it is
required that the fundamental domain for $\Gamma$ is being made of finite
sided polyhedron $P$ in $B^{3}$ (just like for the Riemann surface of finite
genus we should have a finite sided polygon in the unit disk $D$ whose
boundary at infinity is $S_{\infty}^{1}).$ Every hyperbolic manifold M$^{3}$
is defined through use of some fundamental polyhedron $P$ so that, in fact
[42],%
\begin{equation}
M^{3}=(B^{3}\cup\Omega)/\Gamma\tag{5.23}%
\end{equation}
where $\Omega=S_{\infty}^{2}-\Lambda$ is the open set of discontinuity of
$\Gamma$. In general, $\Omega$ represents some collection of Riemann surfaces
which belong to the boundary of $M^{3}.$ This fact has some relevance to
problems associated with 2+1 gravity as explained in Ref.[27],[28]. The
boundary set $\Omega$ is \textbf{not} accessible dynamically however since it
is a $complement$ of the limit set $\Lambda$ in $S_{\infty}^{2}.$ Based in the
information provided, study of the hyperbolic 3-manifolds is equivalent to
study of the action of discrete subroups $\Gamma$ of the M$\ddot{o}$bius group
G on H$^{3}(or$ B$^{3})$. In particular, if the quotient, Eq.(5.23), is
compact, then $\Gamma$ is said to be \textbf{cocompact} and if the quotient,
Eq.(5.23), has finite invariant volume, then $\Gamma$ is said to be
\textbf{cofinite}. Incidentally, if $\Gamma$ contains parabolic subgroups,
then $\Gamma$ is not cocompact. As it was shown by Thurston$[42]($for some
illustrations, please, see also Ref.[70]), complements of most of knots
embedded in $S^{3}$are associated with the hyperbolic 3-manifolds.
Accordingly, if CFT are to be associated with knots/links (e.g.see
Refs[3],[5],[6]), then the corresponding complements of such knots/links, most
likely, should be associated with the hyperbolic 3-manifolds. Moreover, the
spectral characteristics of different hyperbolic manifolds should be different
as well $[19]$. This difference should be also connected with difference in
fractal dimensions of the corresponding limit sets which, in turn, will
correspond to different type (universality classes)of the CFT. Conversely,
given the fractal dimension of the limit set $\Lambda$ , is it possible to
determine the Kleinian \ (or M$\ddot{o}$bius )group (or groups) which is
associated with this limit set ? Evidently, this problem is more complicated
than the direct one. Nevertheless, the above discussion is not limited to
H$^{3}$ ( or B$^{3})$ and, therefore, it becomes potentially possible to study
and to classify boundary CFT in dimensions higher than two. More on this
subject is presented in sections 7 and 8.

Let us now give the precise definitions of parabolic and conical limit points
which were mentioned in the theorem by Beardon and Maskit stated above. An
extensive discussion of both parabolic and conical limit points (and sets)
could be found in Ref.$[71]$. From this reference we find that ''for any
discrete group the set of bounded parabolic points and the set of conical
limit points are disjoint''. Given this, and recalling that the parabolic
transformations are associated with translations we are left with the
following two options(in the case of H$^{3})$ : a)either the parabolic
subgroup has just one generator of translations so that the ''fundamental
polyhedron'' is the region between two parallel planes as depicted in Fig.2.

\mathstrut%
%TCIMACRO{\FRAME{ftbpFU}{2.6325in}{2.4837in}{0pt}{\Qcb{A typical \textbf{Z}%
%-cusp in the upper half space model realization of H$^{3}$}}{}{FIG2.EPS}%
%width 2.6325in;  height 2.4837in;  depth 0pt;  original-width 3.4272in;
%original-height 4.1044in;  cropleft "0";  croptop "1";  cropright "1";
%cropbottom "0.2125";  filename 'FIG2.EPS';file-properties "XNPEU";}}}%
%BeginExpansion
\begin{figure}
[ptb]
\begin{center}
\includegraphics[
trim=0.000000in 0.872185in 0.000000in 0.000000in,
natheight=4.104400in,
natwidth=3.427200in,
height=2.4837in,
width=2.6325in
]%
{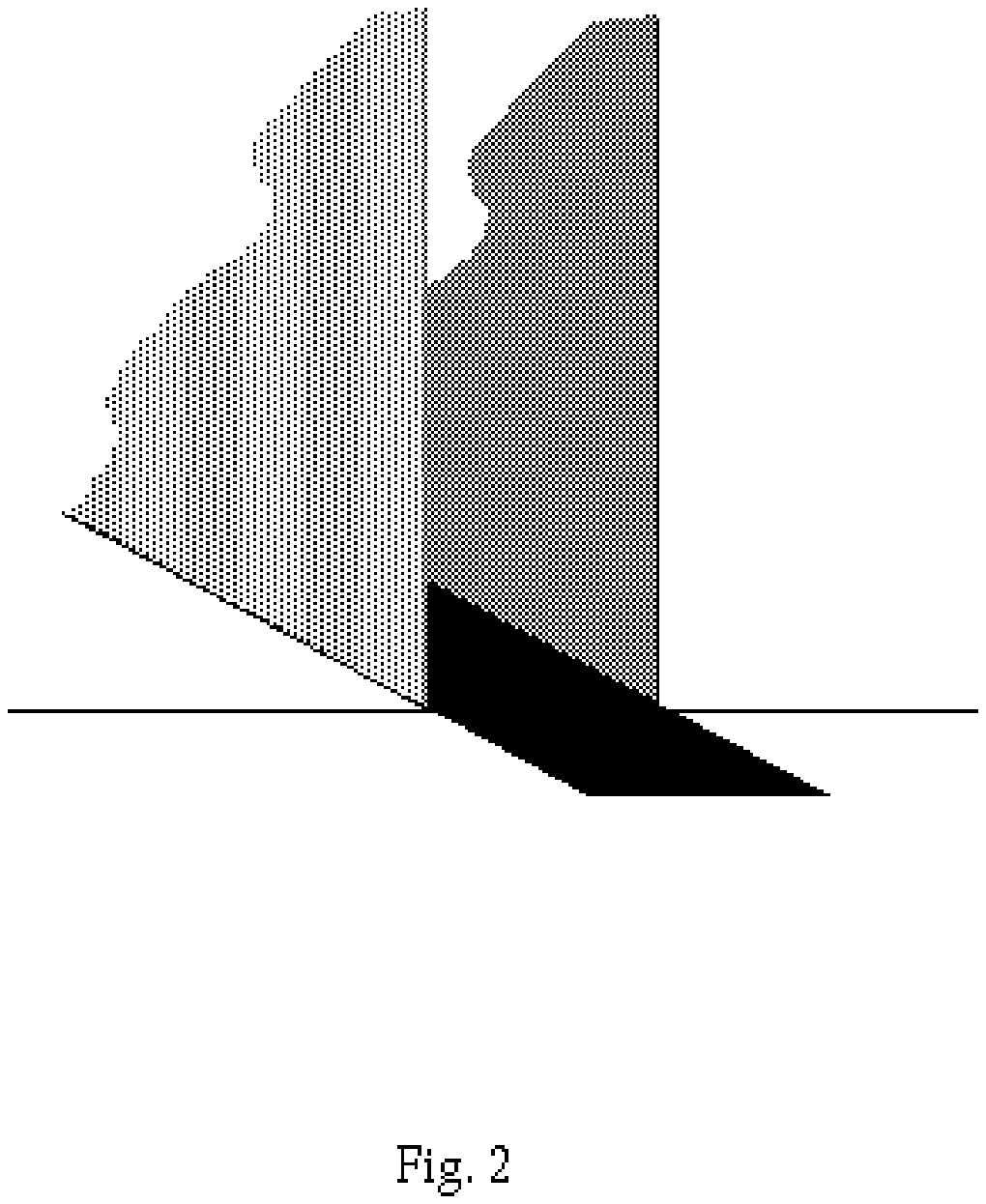}%
\caption{A typical \textbf{Z}-cusp in the upper half space model realization
of H$^{3}$}%
\end{center}
\end{figure}
%EndExpansion

Such construction is called rank 1 ( or \textbf{Z}- cusp).
Evidently,topologically motion $\perp$ to these planes is the same as motion
on the circle $S^{1}$ as it was recently discussed at some length in Ref[70]in
connection with some problems in polymer physics. Accordingly, such parabolic
subgroup is isomorphic to \textbf{Z} or,.b) the parabolic subgroup has two
generators so that the ''fundamental polyhedron'' is the region defined by the
transverse pairs of parallel planes ,as depicted in Fig.3

\mathstrut%
%TCIMACRO{\FRAME{ftbpFU}{2.2087in}{2.2675in}{0pt}{\Qcb{A typical \textbf{Z}%
%$\oplus\mathbf{Z}$ type cusp in the upper half space realization of H$^{3}$}%
%}{}{FIG3.EPS}{\special{ language "Scientific Word";  type "GRAPHIC";
%width 2.2087in;  height 2.2675in;  depth 0pt;  original-width 3.3537in;
%original-height 4.2704in;  cropleft "0";  croptop "1";  cropright "1";
%cropbottom "0.1934";  filename 'FIG3.EPS';file-properties "XNPEU";}}}%
%BeginExpansion
\begin{figure}
[ptb]
\begin{center}
\includegraphics[
trim=0.000000in 0.825895in 0.000000in 0.000000in,
natheight=4.270400in,
natwidth=3.353700in,
height=2.2675in,
width=2.2087in
]%
{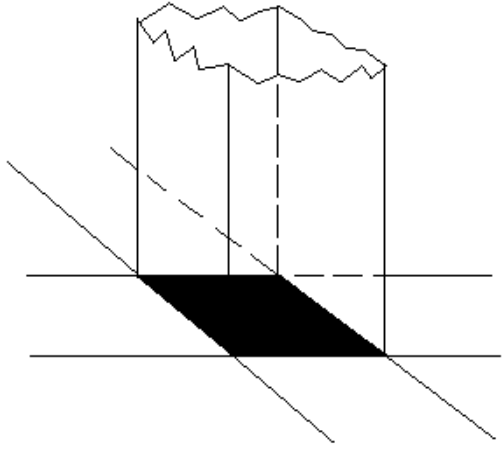}%
\caption{A typical \textbf{Z}$\oplus\mathbf{Z}$ type cusp in the upper half
space realization of H$^{3}$}%
\end{center}
\end{figure}
%EndExpansion

Such construction is called rank 2 ( or \textbf{Z}$\oplus\mathbf{Z}$ ) cusp.
Topologically, motion $\perp$ to such planes is being associated with the
motion on the torus. The restriction to have only \textbf{Z} and
\textbf{Z}$\oplus\mathbf{Z}$ cusps for hyperbolic 3-manifolds imposes very
important restrictions on the boundary CFT to be discussed in section 7.

The conical limit set is not specific to the hyperbolic spaces. According to
Ref.[39], in the case of Euclidean half-space H$_{n}$ for which the typical
point y=(\textbf{x},z), z%
%TCIMACRO{\TEXTsymbol{>}}%
%BeginExpansion
$>$%
%EndExpansion
0, \textbf{x}$\in R^{n-1},$the conical limit set $\Gamma_{\alpha}(a)$ is
defined through
\begin{equation}
\Gamma_{\alpha}(a)=\{(\mathbf{x},z)\in H_{n}:\left|  x-a\right|  <\alpha z\}.
\tag{5.24}%
\end{equation}
Geometrically, $\Gamma_{\alpha}(a)$ is a cone as depicted in Fig.4 with vertex
$a$ and axis of symmetry parallel to z-axis.

\smallskip%
%TCIMACRO{\FRAME{ftbpFU}{3.0684in}{2.9974in}{0pt}{\Qcb{The light cone
%associated with the conical limit set point located at the boundary of
%hyperbolic space}}{}{FIG4.EPS}{\special{ language "Scientific Word";
%type "GRAPHIC";  maintain-aspect-ratio TRUE;  display "USEDEF";
%valid_file "F";  width 3.0684in;  height 2.9974in;  depth 0pt;
%original-width 4.4062in;  original-height 4.8853in;  cropleft "0";
%croptop "1";  cropright "1";  cropbottom "0.1190";
%filename 'FIG4.EPS';file-properties "XNPEU";}}}%
%BeginExpansion
\begin{figure}
[ptb]
\begin{center}
\includegraphics[
trim=0.000000in 0.581351in 0.000000in 0.000000in,
natheight=4.885300in,
natwidth=4.406200in,
height=2.9974in,
width=3.0684in
]%
{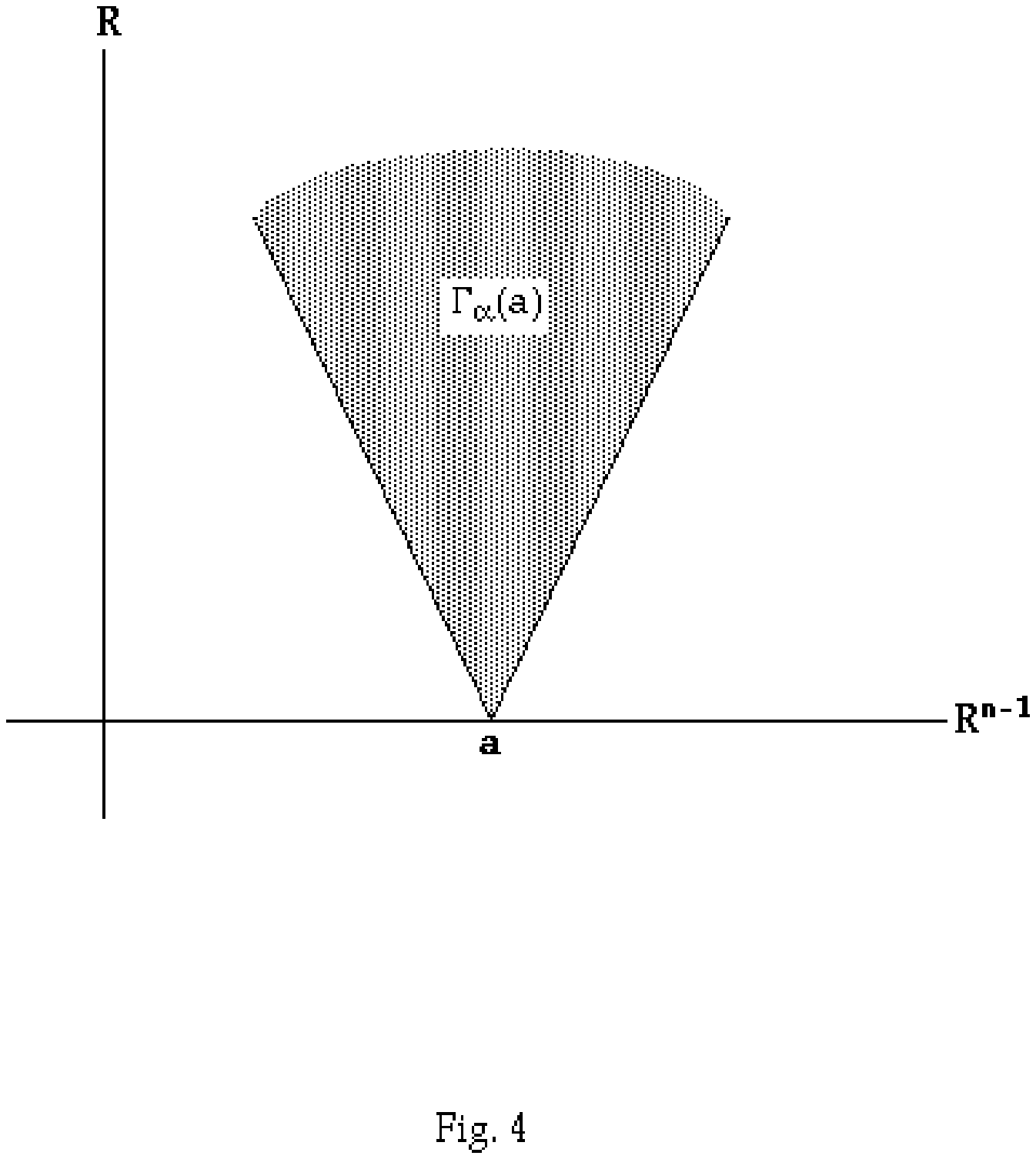}%
\caption{The light cone associated with the conical limit set point located at
the boundary of hyperbolic space}%
\end{center}
\end{figure}
%EndExpansion

A function $u$ defined on H$_{n}$ is said to have \textbf{nontangential limit}
$L$ at $a\in R^{n-1}$ if for every $\alpha>0,u(y)\rightarrow L$ as
$y\rightarrow a$ within $\Gamma_{\alpha}(a).$ The term ''nontangential'' is
being used because no curve in $\Gamma_{\alpha}(a)$ that approaches $a$ can be
tangent to $\partial$H$_{n}=R^{n-1}.$ It is quite remarkable that such
nontangential behavior is being observed already for harmonic functions on
Euclidean half-space H$_{n}[39]$ .Use of stereographic projection allows us to
formulate the same problem in the Euclidean ball B$_{n.}$ Respectively,
exactly the same definitions are extended to H$^{d+1}$ and $B^{d+1}%
.$Specifically$,$ in the case of $B^{d+1}$ model one can say that $x\in
B^{d+1}$ belongs to the cone at $\mathbf{\xi}\in S_{\infty}^{d}$ of opening
$\lambda$ and, further, $\left|  x-\mathbf{\xi}\right|  <2\cos\lambda$ .
Analogously to Eq. (5.24), one can write
\begin{equation}
\left|  x-\mathbf{\xi}\right|  <\alpha(1-\left|  x\right|  ),\text{ }\alpha>0.
\tag{5.25}%
\end{equation}
With such background,we would like discuss in some detail the spectral theory
of hyperbolic 3-manifolds. This is accomplished in the next section.

\mathstrut

{\Large 6. Spectral theory of hyperbolic 3-manifolds}

\mathstrut

In section 3 we had discussed the eigenvalue equation, Eq.(3.7), so that,
naively, one might think that this equation provides the complete answer to
the question about the spectrum of hyperbolic Laplacian . This is not true,
however. Surprisingly, this problem still remains a very active area of
research in mathematics. For a comprehensive and very up to date introduction
to this field, please, consult Ref[28]. The fact that the spectral theory of
hyperbolic Laplacians is absolutely essential for understanding of the
spectrum of Hausdorff dimensions of the limit set $\Lambda$ was realized
already by Patterson[72] long time ago. Since, even now, the spectrum issue is
not completely settled, we would like only to give an outline of the current
situation leaving most of the details for future work.

In his 1987 paper$[70]$ Sullivan had stated the Theorem (2.17)(numeration
taken from his work)which he calls the Patterson-Elstrodt theorem(
incidentally, the recent monograph, Ref[28], \textbf{is} written by Elstrodt
). Based on the results of previous sections it can be formulated as follows:

\textbf{Theorem} \textbf{6.1}.(Patterson-Elstrodt-Sullivan).

\textit{Let }
\begin{equation}
-\Delta_{h}\varphi=\lambda\varphi\tag{6.1}%
\end{equation}
\textit{be the eigenvalue problem for the hyperbolic Laplacian on }%
$M^{d+1}=H^{d+1}/\Gamma$\textit{\ then, the lowest eigenvalue }$\lambda
_{0}(M^{d})$\textit{\ satisfies }
\begin{equation}
\lambda_{0}(M^{d+1})=\left\{
\begin{array}
[c]{c}%
\frac{d^{2}}{4}\text{ if \quad}\alpha\leq\frac{d}{2}\\
\alpha(\Gamma)(\alpha(\Gamma)-d)\text{ if }\alpha\geq\frac{d}{2}%
\end{array}
\right.  \tag{6.2}%
\end{equation}
\textit{where }$\alpha(\Gamma)$\textit{\ is the ''critical'' exponent of the
Poincar}$e^{\prime}$\textit{\ series, Eq.(4.19) or Eq.(5.22)}.

By looking at Eqs.(4.11),(4.12), these results can be restated as
\begin{equation}
m^{2}=\left\{
\begin{array}
[c]{c}%
\lambda_{0}\text{ if }\alpha\geq\frac{d}{2}\\
-\frac{d^{2}}{4}\text{ if }\alpha\leq\frac{d}{2}\text{ .}%
\end{array}
\right.  \tag{6.3}%
\end{equation}
Additional work by Patterson$[62]$ indicates that, at least for $M^{3}$,
$0<\alpha\leq2.$ In view of this , by looking at Eq. (4.12), it is reasonable
to consider both ''+'' and ''-'' branches of solution for $\alpha$ , provided
that $-\frac{d^{2}}{4}<m^{2}\leq0.$ This possibility, indeed, had been
recognized in Ref.[75]. The results obtained by Lax and Phillips$[76]$ (and
also by Epstein$[77])$ indicate that for 3-manifolds \textbf{without parabolic
cusps} the spectrum of -$\Delta_{h}$ acting on $L(M^{3})$ normed metric
Hilbert space is of the form:
\begin{equation}
\{\lambda_{0},...,\lambda_{k}\}\cup\lbrack\frac{d^{2}}{4},\infty) \tag{6.4}%
\end{equation}
where
\begin{equation}
0<\alpha(d-\alpha)=\lambda_{0}<\lambda_{1}<...<\lambda_{k}<\frac{d^{2}}{4}
\tag{6.5}%
\end{equation}
are eigenvalues of finite multiplicity and $\lambda_{0}$ has multiplicity one.
Moreover, the part of spectrum $[\frac{d^{2}}{4},\infty)$ is absolutely
continuous (i.e.for $m^{2}\leq-\frac{d^{2}}{4}$ the spectrum \textbf{is}
continuous). The problem with Lax-Phillips$[76]$ and Epstein$[77]$ works lies,
however, in the fact that the \textbf{explicit} form of the discrete spectrum
had not been obtained. Only the existence of such possibility had been proven.

\textbf{Remark} \textbf{6.2}. In view of Beardon-Maskit Theorem (section 5)
one cannot by pass careful study of the spectrum of hyperbolic Laplacian for
some discrete subgroups $\Gamma$ of M$\ddot{o}$bius group G if one is
interested in finding the correct fractal dimension of the limit set $\Lambda$ .

For the sake of applications to statistical mechanics (e.g.see section 7) one
is also interested in spectral properties of 3-manifolds with parabolic cusps.
This can be intuitively understood already now based on the following
arguments. If we would choose the sign ''-'' in Eq. (4.12) (which,by the way
would produce ''+'' sign in front of Eq. (4.16), then for $m^{2}$ in the range
$-\frac{d^{2}}{4}\leq m^{2}<0$ we would have $\alpha$ in the range
$0<\alpha\leq1$ for d=2. This range is of interest since it covers all
physically interesting CFT discussed in the literature$[9].$ If $\alpha$ is to
be associated with the Hausdorff dimension of the limit set $\Lambda$, then
according to Sullivan ,e.g.see Theorem 2 of Ref.[66], only 3 manifolds with no
cusps or rank 1 (Fig.2) cusps will yield $\alpha$ in the desired range. The
spectral theory of hyperbolic manifolds with cusps is still under active
development in mathematics[29]. Therefore, we would like to restrict ourself
with some qualitative estimates of the spectrum based on topological
arguments. Here and below we shall discuss only the case d=2 (i.e.H$^{3}$ or
B$^{3})$. This restriction is by no means severe. It is motivated only by the
fact that more explicit analytical results are available for this case in
mathematical literature. This, however, does \textbf{not} imply that the case
d=2 is more special than say d=3. For instance, Burger and Canary $[78]$ had
demonstrated that for any d$>$1 the Hausdorff dimension $\alpha$ is bounded
by
\begin{equation}
\alpha\leq(d-1)-\frac{K_{d}}{(d-1)vol(C(M^{d}))^{2}} \tag{6.6}%
\end{equation}
where $K_{d}$ and $C(M^{d})$ are some d-dependent constants which can be
calculated in principle .

In the case if hyperbolic manifold $M^{3}$ is topologically \textbf{tame}
(that is it is homeomorphic to the interior of a compact 3 manifold), then
Theorem 2.1. of Canary et all $[79]$ states that

\textbf{Theorem} \textbf{6.3}.(Canary, Minsky and Taylor) \textit{If M}$^{3}%
$\textit{\ is topologically tame hyperbolic 3-manifold, then the lowest
eigenvalue }$\lambda_{0}$\textit{\ of the hyperbolic Laplacian (-}$\Delta
_{h})$\textit{\ is given by }$\lambda_{0}=\alpha(2-\alpha)$\textit{\ unless
}$\alpha<1,$ \textit{in which case }$\lambda_{0}(M^{3})=1.$

\textbf{Remark} \textbf{6.4} As before, $\alpha$ is the Hausdorff dimension of
the limit set $\Lambda.$

\textbf{Remark} \textbf{6.5} From the above Theorem 6.3. it appears that the
results of section 4 become invalid when $\alpha<1$ since Eq.(4.12) cannot be
used. The situation can be easily repaired as it is explained in the next section.

\textbf{Remark} \textbf{6.6}. Theorem 6.3.allows us to obtain the following
additional estimates based on recent results by Bishop and Jones$[43].$

\textbf{Theorem} \textbf{6.7}.(Bishop and Jones). \textit{Let }$\Gamma
$\textit{\ be any discrete M}$\ddot{o}$\textit{bius group and let }%
$M^{3}=(B^{3}\cup\Omega)/\Gamma$\textit{\ . Suppose that the lowest eigenvalue
}$\lambda_{0}$\textit{\ is nonzero.Then, there are \ constants }$C<\infty
$\textit{\ and }$c>0$\textit{\ (depending upon }$\lambda_{0}$\textit{\ only)so
that for any x,y with }$\rho(\mathbf{x},\mathbf{y})\geq8$\textit{\ we have }
\begin{equation}
G(\mathbf{x},\mathbf{y})=\int\limits_{0}^{\infty}dtG(\mathbf{x},\mathbf{y}%
;t)\leq\frac{C}{\lambda_{0}}\exp\{-c\rho(\mathbf{x},\mathbf{y})\} \tag{6.7}%
\end{equation}
\textit{where }$\rho(\mathbf{x},\mathbf{y})$\textit{\ is the hyperbolic
distance between x and y and }$c$\textit{=}$\min\{\frac{1}{8}\lambda_{0}%
,\frac{1}{4}\}.$

\textbf{Corollary}.\textbf{6.8}.Using this result in combination with
Eqs.(4.18),(4.19) and the Theorem 6.3. we obtain, $\alpha=\frac{\lambda_{0}%
}{8}=\frac{1}{8}.$ If this result is substituted into Eq.(4.16) we obtain the
exact result for two-point correlation function of two dimensional Ising model.

\textbf{Remark} \textbf{6.9}.The theorem of Bishop and Jones depends crucially
on the explicit form for the heat kernel $G(\mathbf{x},\mathbf{y};t)$ in
H$^{3}$. Quite recently, Grigori'yan and Noguchi$[80]$ had obtained explicit
formulas for the heat kernel for any dimension of hyperbolic space. This opens
a possibility to obtain an analogue of inequality (6.7) in any dimension
following ideas of Bishop and Jones.

With all plausibility of the Corollary 6.8. it remains to demonstrate that
such substitution of $\alpha$ into Eq. (4.16) is indeed legitimate.To this
purpose we would like to provide a somewhat different interpretation of
Eq.(4.16) in order to demonstrate that Eq.(4.17) makes sense even without
arguments associated with Plateau/Dirichlet problem. To begin, we would like,
by analogy with the Liouville theorem in standard textbooks on statistical
mechanics, to construct a measure associated with the geodesic flow in
hyperbolic space.

Following Ref.[25] ,we would like to associate with each point $x\in
B^{d+1}\equiv B$ a unit vector $\mathbf{\xi}\in S_{{}}^{d}$of directions. This
vector plays the same role as velocity $\mathbf{v}$ in conventional
statistical mechanics. Indeed,$\forall\mathbf{v\neq0}$ one can construct a
vector $\mathbf{\xi}$=$\frac{\mathbf{v}}{\left|  \mathbf{v}\right|  }$ and
then proceed with standard development. The M$\ddot{o}$bius group $\Gamma$ is
acting on the phase space T(B)=B$\times S$ according to the rule
\begin{equation}
\left(  x,\xi\right)  \rightarrow\left(  \gamma x,\frac{\gamma^{^{\prime}}%
(x)}{\left|  \gamma^{^{\prime}}(x)\right|  }\xi\right)  ,\text{ }\forall
\gamma\in\Gamma. \tag{6.8}%
\end{equation}
The invariant phase space volume element $d\Omega$ is given therefore by
\begin{equation}
d\Omega=dx_{h}d\omega(\xi) \tag{6.9}%
\end{equation}
with $d\omega(\xi)$ being a spatial angle measure and dx$_{h}$ being an
element of a hyperbolic volume. The above chosen variables may not be the most
convenient ones. More convenient are variables associated with actual location
of the ends of geodesics $u$ and $v$ on $S_{\infty}^{d}.$This situation is
depicted in Fig.5

\mathstrut%
%TCIMACRO{\FRAME{ftbpFU}{2.1032in}{2.7259in}{0pt}{\Qcb{Geometry of geodesics in
%the ball model of hyperbolic space}}{}{FIG5.EPS}%
%width 2.1032in;  height 2.7259in;  depth 0pt;  original-width 3.0519in;
%original-height 4.8646in;  cropleft "0";  croptop "1";  cropright "1";
%cropbottom "0.1843";  filename 'FIG5.EPS';file-properties "XNPEU";}}}%
%BeginExpansion
\begin{figure}
[ptb]
\begin{center}
\includegraphics[
trim=0.000000in 0.896546in 0.000000in 0.000000in,
natheight=4.864600in,
natwidth=3.051900in,
height=2.7259in,
width=2.1032in
]%
{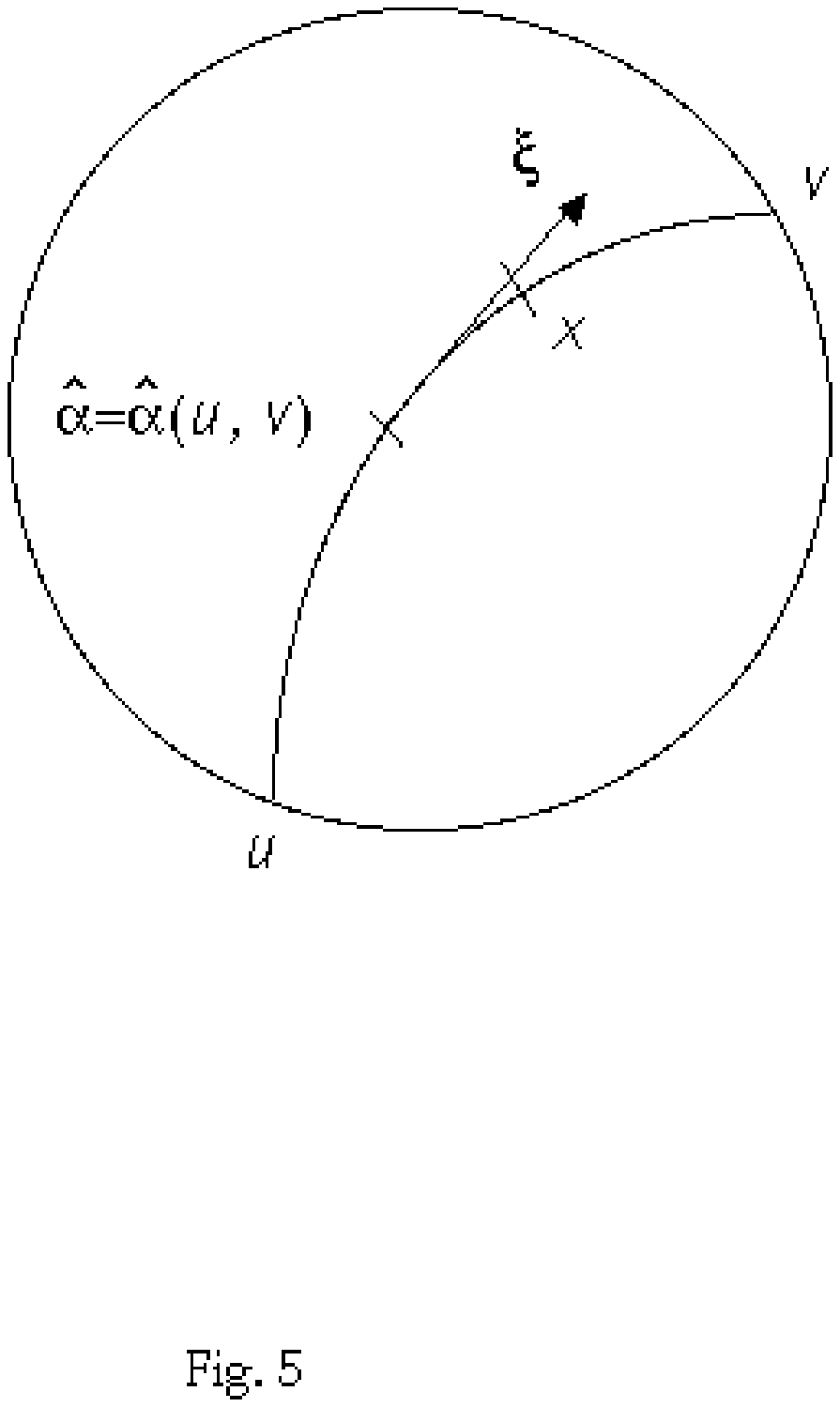}%
\caption{Geometry of geodesics in the ball model of hyperbolic space}%
\end{center}
\end{figure}
%EndExpansion

It is clear, that $\forall x\in B$ one can select a geodesic which passes
through $x$. To this purpose it is not sufficient to assign $u$ and $v$ on
$S_{\infty}^{d}$ but, in addition, one has to provide a location $\hat{\alpha
}(u,v)$ of the midpoint for such geodesics. Let $s$ be the directional
hyperbolic distance between $\hat{\alpha}$ and $x$, then, one should be able
to find a correspondence between ($x$,$\xi)$ and $(u,v,s),$ that is one should
be able to find a diffeomorphism between $B\times S$ and $S\times S\times R,$
i.e.one expects to find an explicit form of the function $f(u,v)$ which enters
into the expression for the volume element given below:
\begin{equation}
d\Omega=dx_{h}d\omega(\xi)=f(u,v)d\omega(v)d\omega(u)ds \tag{6.10}%
\end{equation}
A simple argument given in Ref[25] produces
\begin{equation}
f(u,v)=\frac{G}{\left|  u-v\right|  ^{2d}} \tag{6.11}%
\end{equation}
with $G$ being some normalization constant. Looking now at Eq. (3.22), it is
clear, that one can now replace it with
\begin{equation}
\hat{D}[\varphi]=\int\frac{d\Omega}{ds}\varphi_{0}(u)\varphi_{0}(v)\text{ .}
\tag{6.12}%
\end{equation}
It is also clear, in view of the transformation properties of the function
$\varphi_{0}$given by Eq.(4.14), that, in general, one can replace Eq.(6.12)
with
\begin{equation}
\hat{D}[\varphi]=G\int\frac{\varphi_{0}(u)\varphi_{0}(v)}{\left|  u-v\right|
^{2\alpha}}d\omega(v)d\omega(u) \tag{6.13}%
\end{equation}
where the exponent $\alpha$ is associated with the Hausdorff dimension of the
limit set $\Lambda$ .This is indeed the case, e.g.see page 286 of Ref[64].
Thus, the exponent $\alpha$ in Eq. (6.13) is\textbf{\ the same} as the
exponent $\alpha$ in Eq. (5.22). This observation provides the necessary
support to the claims made after Eq.(6.7).

Given all above, the obtained results show no apparent connections with the
existing conformal field theories. We would like to correct this deficiency in
the next two sections.

\mathstrut

{\Large 7. Connections with the existing formalism of CFT}

\mathstrut

In section 5 we had introduced \textbf{Z} and \textbf{Z}$\oplus\mathbf{Z}$
cusps, e.g. see Figs 2,3. According to Sullivan$[66],$only 3-manifolds with no
cusps or just \textbf{Z}-cusps will produce limit sets $\Lambda$ with
Hausdorff dimension $\alpha$ in the range $0<\alpha\leq1.$ Naively, it means,
that only consideration of the CFT on the strip with periodic boundary
conditions(thus making it a cylinder) will yield the critical exponents for
two point correlation functions in the above range. This case is, indeed,
frequently discussed in physics literature$[9].$ For the strip of width L use
of the conformal transformation
\begin{equation}
z^{\prime}=w(z)=\frac{L}{2\pi}\ln z \tag{7.1}%
\end{equation}
converting strip of width $L$ to the entire complex plane (rigorously
speaking, we are dealing here with $\mathbf{C\backslash\{0\}}$ complex
plane$[81]$). Although the above discussion appears to be plausible ,the
description of \textbf{Z}-cusps (as well as \textbf{Z}$\oplus\mathbf{Z}$ cusps
) is actually considerably more sophisticated. In this paper we only provide a
brief outline of what is actually involved reserving full treatment for future publications.

In section 5 we had noticed that $\Lambda$ may contain no more than two
limiting points (elementary set) or infinite number of points (non-elementary
set). The Kleinian groups which are associated with the elementary limit sets
are known $[8],[31]$ and, basically, are reducible to the following list:

1) a parabolic infinite cyclic Abelian group $\Gamma:$ $z\rightarrow z+1;$

2) a parabolic rank 2 Abelian group $\Gamma$ : $z\rightarrow z+1,z\rightarrow
z+\tau$ ; $\operatorname{Im}\tau>0;$

3) a loxodromic cyclic group $\Gamma$ : z$\rightarrow\lambda z$ with
$\lambda\in\mathbf{C}\backslash\{0,1\}.$

Let now $M^{3}$ be some 3-manifold and let $M_{(0,\varepsilon)}$ be a subset
of points $p\in M^{3}$ such that there is a closed nontrivial curve passing
through $p$ whose hyperbolic length $l$ is less than $\varepsilon$. Then, if
$\varepsilon<2r_{0}$ where $r_{0}$ is some known (Margulis)constant, the
$M_{(0,\varepsilon)}$ part of $M^{3}$ (the ''thin part'') is a quotient
$H^{3}/\Gamma$ where $\Gamma$ is just one of these three elementary groups.
The complement of $M_{(0,\varepsilon)}$ in $M^{3}$ is called ''thick'' part.
The above construction is not limited to $M^{3}$and is applicable to
$\mathbf{any}$ $M^{d+1}($with Margulis constant being, of course, different
for different d's). The ''thin'' part is associated with \textbf{Z }and
\textbf{Z}$\oplus\mathbf{Z}$ cusps.

\textbf{Remark 7.1}.Recently, we had briefly considered the ''thick ''
-''thin'' decomposition of hyperbolic 3-manifolds in connection with dynamics
of 2+1 gravity $[36],[37].$ For a comprehensive mathematical treatment of
these issues, please, consult Ref[42],[71],[82].

To realize, that the ''thin'' part is associated with \textbf{Z} cusps it is
sufficient to look at H$^{2}$ model of hyperbolic space first. In this case,
the following Theorem can be proven$[83]$

\textbf{Theorem 7.2}. \textit{Let G be Fuchsian group operating on H}$^{2}%
$\textit{\ .If G contains a parabolic element, then H}$^{2}/G$%
\textit{\ contains a puncture.The number of punctures is in one-to-one
correspondence with the number of conjugacy classes of parabolic elements.}

Recall now, that in 3 dimensional case $\Omega=S_{\infty}^{2}-\Lambda$ and,
using Eq.(5.27), it is possible to show that $\Omega/\Gamma$ is just a
collection of Riemann surfaces $[44]$ . In the case if we are dealing with
Z-cusps these surfaces will contain punctures as it was first noticed by
Ahlfors$[84]$. The number of cusps (=punctures) N$_{c}$ is related to the
number of generators N of the Kleinian group acting on H$^{3}.$ According to
Sullivan$[85]$(and also Abikoff$[86])$%
\begin{equation}
\text{N}_{c}\leq3\text{N-4} \tag{7.2}%
\end{equation}
In the language of the CFT the punctures are usually associated with the
vertex operators$[9].$ The presence of punctures converts Riemann surface
$R$=$\Omega/\Gamma$ into the \textbf{marked} Riemann surface$[27].$ We shall,
for simplicity, treat the quotient $\Omega/\Gamma$ as just \textbf{one}
Riemann surface (unless the otherwise is specified) keeping in mind that there
could be\textbf{\ finitely} many(Ahlfors finiteness theorem$[84])$. Among
marked surfaces one can choose some reference Riemann surface X for which the
marking is fixed. Then, \textbf{other} surfaces could be related to X via
homeomorphism f: $R\rightarrow$X sending the orientation on R into orientation
on X. The \textbf{Teichm\"{u}ller space}, Teich($R$), associates conformal
structures on R in which each boundary component corresponds to a puncture.
Two marked surfaces (f$_{1,}R_{1})$ and (f$_{2},R_{2})$ define \textbf{the
same } point in Teichm\"{u}ller space Teich(R) if there is a complex analytic
isomorphism i: $R_{1}\rightarrow R_{2}$ such that i$\circ$f$_{1}$ is homotopic
to f$_{2}.$ Two surfaces $R_{1}$ and $R_{2}$ belong to two different points in
Teichmuller space if the Teichm\"{u}ller metric (distance)
\begin{equation}
d(R_{1},R_{2})=\frac{1}{2}\inf\ln K(\phi) \tag{7.3}%
\end{equation}
is greater than zero. Here $\phi:$ $R_{1}\rightarrow R_{2}$ ranges over all
quasiconformal maps in the homotopy class f$_{2}\circ$f$_{1}^{-1}($relative to
the punctures) so that K($\phi)$ is \textbf{maximum dilatation} of $\phi.$ The
above formula is not immediately useful since we have not defined yet what is
meant by dilatation.To correct this deficiency, let us consider the Beltrami
coefficient (for suggestive physical interpretation, please, consult Ref[37])
\begin{equation}
\mu_{f}=\frac{\partial_{\bar{z}}f(z)}{\partial_{z}f(z)} \tag{7.4}%
\end{equation}
For functions f$_{1}$and f$_{2}$ introduced above we obtain respectively
$\mu_{1}$ and $\mu_{2}.$Then,the maximum dilatation can be defined as
\begin{equation}
K(\phi)=\frac{1+r}{1-r}\text{ \quad where }r\text{=}\left\|  \frac{\mu_{1}%
-\mu_{2}}{1-\bar{\mu}_{1}\mu_{2}}\right\|  _{\infty} \tag{7.5}%
\end{equation}
according to [30],[44], with $\left\|  \cdot\cdot\cdot\right\|  _{\infty}$
being determined by the requirement $[44]$%
\begin{equation}
\left\|  \mu_{f}(z)\right\|  =\sup_{z\in R}\left|  \mu_{f}(z)\right|  <1.
\tag{7.6}%
\end{equation}
From the above results it follows, that if $\gamma\in\Gamma$ and z$\in
S_{\infty}^{2},$ then
\begin{equation}
\mu(\gamma(z))\frac{\bar{\gamma}^{\prime}(z)}{\gamma(z)}=\mu(z)\text{ \quad
}\forall z\in\Omega\tag{7.7}%
\end{equation}
and
\begin{equation}
\mu(z)=0\text{ \quad}\forall z\in\Lambda. \tag{7.8}%
\end{equation}
Let us now fix $\mu$ and introduce f$^{\mu}(z)$ instead (that is f$^{\mu}(z)$
is some function which produces the Beltrami coefficient according to
Eq.(7.4)). The mapping $\Gamma\rightarrow\Gamma^{\mu}$ given by $\gamma
\rightarrow$f$^{\mu}\circ\gamma\circ($f$^{\mu})^{-1}$ is called
\textbf{quasiconformal} (or $\mu-conformal)$ deformation. Let us notice now
that normally the Riemann surface $R$ is being defined as quotient $R$%
=H$^{2}/G$ where G is some discrete Fuchsian group. In the case of
$S_{\infty\text{ }}^{2}$we have a rather peculiar situation: Kleinian group
$\Gamma\subset PSL_{2}(C)$ plays \textbf{the same} role as Fuchsian G$\subset
PSL_{2}(R).$ One can bring these two together by noticing that H$^{2}$ model
corresponds to an open disc D. Then, one can glue two copies of D together
thus forming $S_{\infty\text{ }}^{2}.$ Kleinian group $\Gamma$ acting on
$S_{\infty}^{2}$ can be considered as Fuchsian on each of these two disks. The
mapping $\Gamma\rightarrow\Gamma^{\mu}$ may affect the gluing boundary between
the two disks. If we use f$^{\mu}$ to produce ''new'' group from the ''old'',
i.e.,
\begin{equation}
\gamma^{\mu}=f^{\mu}\circ\hat{\gamma}\circ(f^{\mu})^{-1}, \tag{7.9}%
\end{equation}
then thus obtained new group is called \textbf{quasi-Fuchsian} (provided that
$\hat{\gamma}$ is Fuchsian) if the gluing boundary between two disks is still
topologically a circle $S^{1}(e.g.$ see Thurston's lecture notes$[42]$,
section 8.34). This gluing boundary may include $\Lambda$ as a part only or,
it could be that $\Lambda=S_{{}}^{1}.$

Recently, Canary and Taylor $[46]$ had proved the following remarkable theorem.

\textbf{Theorem} \textbf{7.3}.(Canary and Taylor). \textit{Let }$\Gamma
$\textit{\ be a nonelementary finitely generated Kleinian group and let
}$\Lambda$\textit{\ denote its limit set. If the Hausdorff dimension }$\alpha
$\textit{\ of }$\Lambda$\textit{\ is less than one, then }$\Gamma$\textit{\ is
geometrically finite and has a finite index subgroup which is quasiconformally
conjugate to a Fuchsian group of the second kind .}

\textbf{Remark 7.4}. Recall$[87]$, that for the Fuchsian groups of the
\textbf{second} kind the limit points are nowhere dense on $S^{1}.$Since,
according to results of section 6, we are interested mainly in the $\alpha
-$domain given by $0<\alpha\leq1,$ we notice that we have to deal with the
quasiconformal deformations of $S^{1}$ associated with Fuchsian groups of the
second kind.

For completeness, we would like also to provide the results related mainly to
the Fuchsian groups of the \textbf{first} kind for which the limit set
$\Lambda$ coincides with $S.$ These are summarized in the following

\textbf{Theorem} \textbf{7.5}.(Canary and Taylor$[46])$ \textit{Let }$\Gamma
$\textit{\ be a nonelementary finitely generated Kleinian group and let
}$\Lambda$\textit{ denote its limit set}. If $\alpha=1$\textit{, then }%
$\Gamma$\textit{\ either a function group with connected domain of
discontinuity or contains a subgroup of index at most 2 which is the Fucsian
group of the \textbf{first} kind. Alternatively, if \ }$\alpha=1$ \textit{and
}$\Gamma$\textit{ is geometrically finite, then either }$\Gamma$\textit{ has a
finite index subgroup which is quasiconformally conjugate to a Fuchsian}
\ \textit{group of the \textbf{second} \ kind or }$\Gamma$\textit{ contains a
subgroup of index at most 2 which is the Fuchsian group of the \textbf{first} kind.}

\textbf{Remark} \textbf{7.6}.Much earlier Bowen$[88]$ had proven an analogous
theorem for the Fuchsian groups of the first kind. According to Bowen, the
Hausdorff dimension of $\Lambda$ is greater than one. Since Bowen's proof is
nonconstructive, there is no way to estimate, based on his results, to what
extent $\alpha$ is larger than one. Thus, there is no contradiction between
Theorem 7.5. and Bowen's results since $\alpha$ can be infinitesimally close
to 1.

\textbf{Remark} \textbf{7.7}.For the case of Fuchsian groups of the first kind
it is known $[88],$ that $\Omega/\Gamma$ consists of \textbf{exactly} two
Riemann surfaces :one for each disk D. It is also known$[45],$ that for the
Fuchsian group of the\textbf{\ second }kind $\Omega/\Gamma$ is made of just
\textbf{one} Riemann surface so that $S_{\infty}^{2}$ is boundary at infinity
for this surface.

In mathematics literature$[31]$ a finitely generated non elementary Kleinian
group which has just one invariant component $\Omega$ is called
\textbf{function} group. If, in addition, $\Omega$ is simply connected, then
such group is called B-group. More complicated Kleinian groups could be
constructed from simpler ones and B-group is one of the main building blocks
in such construction[90].

\textbf{Remark} \textbf{7.8}.In string theory (and, therefore, in the CFT)the
Schottky-type groups are being used$[91]$. Schottky group is a function group
but \textbf{not} a B-group [31].

\textbf{Remark} \textbf{7.9.} There is one-to one correspondence between the
quasiconformal deformations of Kleinian groups and quasisometric deformations
of hyperbolic 3-manifolds$.$ The theory is not limited to 3-manifolds,
however, and can be considered for any d$\geq2.$ More specifically, there is a

\textbf{Theorem} \textbf{7.10}. \textit{For a quasiconformal automorphism f of
}$S_{\infty}^{2}$\textit{\ compatible with a Kleinian group }$\Gamma$\textit{,
there exist a quasi-isometric automorphism F of H}$^{3}$\textit{\ which is an
extension of f and which is compatible with }$\Gamma$\textit{, namely,
F}$\circ\gamma\circ$\textit{F}$\in Mob(B^{3})$\textit{\ for any }$\gamma
\in\Gamma.$

\textbf{Proof}: Please, consult Ref.[31] (page 157).$\square$

\textbf{Remark 7.11}.The above theorem follows directly from the discussion
related to Eq.s (7.7)--(7.9) and for additional details and motivations,
please, consult the work by Bers, Ref.[45]$.$

\textbf{Remark} \textbf{7.12}.The above Theorem is applicable to the case
when, instead of $S_{\infty}^{2},$ we use $S_{\infty}^{1}($taking into account
the results on Canary and Taylor, Theorems 7.3 and 7.5)

The observations presented above allow us to make a \textbf{direct connection}
with the existing results associated with 2 dimensional CFT. To begin, let us
notice that if we would have $S_{\infty}^{1}$ as limit set $\Lambda$ for the
Fuchsian groups of the first kind, then, according to Eq. (7.8) we
could\textbf{\ not} use the quasiconformal mapping and, accordingly,we would
be stuck with just one conformal structure. This fact is known in mathematics
as Mostow rigidity theorem. Usually, this theorem is applied to spaces of
dimensionality $\geq3$ (for more details,please,see section 8).At the same
time, if we consider Fuchsian groups of the second kind, then, we need to deal
with maps of $S_{\infty}^{1}$ acting on some \textbf{open} intervals (since
$\Lambda$ is \textbf{closed }set) of $S_{\infty}^{1}.$ This is not exactly the
situation which is known in physics literature. Indeed, in physics literature
on CFT one is dealing with the Virasoro algebra. Let us recall how one can
arrive at this algebra. Following Ref.[56], let us consider the group
G=$DiffS^{1}$ of orientation preserving diffeomorphisms of $S_{\infty}^{1}.$
Let $\alpha_{1}(z)$ and $\alpha_{2}(z)$ be two elements of G, then the group
composition law can be defined by
\begin{equation}
\alpha_{1}\circ\alpha_{2}(z)=\alpha_{1}(\alpha_{2}(z))\text{ \quad, }%
z=\exp\{i\theta\}. \tag{7.10}%
\end{equation}
The representation of the group G is defined according to the following
prescription:
\begin{equation}
U(\alpha)f(z)=f(\alpha^{-1}(z)), \tag{7.11}%
\end{equation}
where the operator $U(\alpha)$ acts on the vector space of \textbf{smooth}
complex-valued functions on $S_{\infty}^{1}.$The explicit form of the operator
$U(\alpha)$ can be easily found if one notices that
\begin{equation}
\alpha(z)=z(1+\varepsilon(z))=z+\sum\limits_{n=-\infty}^{\infty}%
\varepsilon_{n}z^{n+1},\text{ }\varepsilon_{n}\rightarrow0^{+}. \tag{7.12}%
\end{equation}
Using this expansion and keeping only terms up to 1st order in $\varepsilon
_{n}$ we obtain,
\begin{equation}
U(\alpha)f(z)=f(z-\sum\limits_{n=-\infty}^{\infty}\varepsilon_{n}%
z^{n+1})=(1+\sum_{n}\varepsilon_{n}\hat{d}_{n})f(z) \tag{7.13}%
\end{equation}
with operator $\hat{d}_{n}^{{}}$ given by
\begin{equation}
\hat{d}_{n}^{{}}=-z^{n+1}\frac{d}{dz}=i\exp\{in\theta\}\frac{d}{d\theta}.
\tag{7.14}%
\end{equation}
The operators $\hat{d}_{n}^{{}}$ form a closed Lie algebra $VectS^{1}$
described in terms of the following commutator
\begin{equation}
\left[  \hat{d}_{m},\hat{d}_{n}^{{}}\right]  =(m-n)\hat{d}_{m-n}^{{}}
\tag{7.15}%
\end{equation}
The central extension of this algebra (to be discussed later in this section)
produces the Virasoro algebra. $VectS^{1}$ contains a closed subalgebra formed
by $\hat{d}_{0}^{{}},\hat{d}_{1}^{{}}$ and $\hat{d}_{-1}$ corresponding to the
infinitesimal conformal transformations of the extended complex plane $S^{2}%
$=$\mathbf{C\cup\{\infty\}}$ caused by the action of PSL$(2,C).$ Thus, even
though we had started with diffeomorphisms of the circle, we ended up with the
automorphisms of the extended complex plane. The question arises: is such
extension unique ? The answer is: ''no'' ! Because of this negative answer,
there is a real possibility of extension of the operator formalism of 2d CFT
to higher dimensions. This issue is going to be discussed in the next section.
For the time being,we would like to explain the reasons why the answer is ''no''.

Following Ahlfors$[92]$, and, more recently, Gardiner and Sullivan$[93],$we
would like to consider a quasisymmetric mapping (to be defined below) of the
disk D to itself which induces a topological mapping of the circumference,
i.e. S$_{\infty}^{1}.$ To this purpose it is convenient to use a
\textbf{conformal} transformation which converts the disk model to the upper
halfplane Poincar$e^{\prime}$ model of the hyperbolic space H$^{2}.$ Next, it
is convenient to select points x,x-t, and x+t on the real line \textbf{R}
(corresponding to $S_{\infty}^{1})$ so that the mapping $h(x)$ satisfies the
M-condition
\begin{equation}
M^{-1}\leq\frac{h(x+t)-h(t)}{h(x)-h(x-t)}\leq M \tag{7.16}%
\end{equation}
Let $h$ be a homeomorphism mapping of an open intrerval I of the real axis
into the real axis. Then, $h$ is \textbf{quasisymmetric} on I if there exist a
constant M such that the inequality (7.16) is satisfied for all x-t, x, x+t in
I. Thus defined quasisymmetric mapping forms a group(which we shall denote as
QS) which obeys \textbf{the same} composition law as given by Eq.(7.10)
(except now z is on the real line). The real line $\mathbf{R}$ is the
universal covering of the circle. The exponential mapping, exp (2$\pi
i\theta),$ induces an isomorphism between \textbf{R}/\textbf{Z} and $S^{1}.$
The homeomorphism h(x) of $S_{{}}^{1}$which is characterized by the properties
$[93]$%
\[
h(0)=0,\text{ }h(x)+1=h(x+1)
\]
can be lifted to a homeomorphism $\tilde{h}(x)$ of $\mathbf{R}$ which obeys
the following inequalities:
\begin{equation}
1-\varepsilon(t)\leq\frac{\tilde{h}(x+t)-\tilde{h}(x)}{\tilde{h}(x)-\tilde
{h}(x-t)}\leq1+\varepsilon(t), \tag{7.17}%
\end{equation}
where $\varepsilon$ converges to zero with t.

It is easy to check that this result is consistent with Eq.(7.12) and,
therefore, the group G=$DiffS^{1}$ is called the group of symmetric
homeomorphisms. G is proper subgroup of QS$[93].$ Looking at Eq.(7.5) and
identifying $r$ with $\varepsilon(t)$ we conclude, that the subgroup G has
boundary dilatation asymptotically equal to one. That is such transformation
\textbf{do not} cause the deformations of hyperbolic 3-manifolds. The above
deficiency of the group G was recognized and corrected in the fundamental work
by Nag and Verjovsky $[48].$ Below, we would like to provide the summary of
their accomplishments in the light of results just described and with purpose
of extension of these results in section 8. In order to do so, we still need
to make several observations related to QS group. Let us begin with

\textbf{Theorem} \textbf{7.13}.(Ahlfors-Beurling$[94])$ \textit{Assume }%
$h$\textit{\ is homeomorphism of R .Then, h is quasisymmetric if, and only if,
there exists a quasiconformal extension }$\tilde{h}$\textit{\ of }%
$h$\textit{\ to the complex plane . If }$h$\textit{\ is normalized to fix
three points, say 0,1 and }$\infty,$ \textit{then }$h$\textit{\ is
quasisymmetric with constant M. The quasiconformal extension }$\tilde{h}%
$\textit{\ can be selected so that its dilatation }$K$\textit{\ is less than
or equal to }$c_{1}(M)$\textit{\ where }$c_{1}(M)\rightarrow1$\textit{\ as
}$M\rightarrow1.$

\textbf{Remark} \textbf{7.14.} The symmetric homeomorphism $\alpha(z)$ by
contrast fixes only one point: z=0. Some explicit examples of construction of
$\tilde{h}$ are given in the papes by Carleson$[95]$ and Agard and Kelingos [96].

\textbf{Remark} \textbf{7.15}.Because $c_{1}(M)\rightarrow1$ when
M$\rightarrow1$ any symmetric homeomorphism of S$^{1}$ can be approximated by
quasisymmetric one. \textit{This is the most important fact facilitating
development of conformal field theories beyond 2 dimensions.}

\textbf{Remark 7.16.} Construction of $h$ is closely related to study of maps
of the circle as it is known in the theory of dynamical systems$[97]$.
Evidently,one is interested in maps which map points $\in\Lambda$ to points in
$\Lambda$ and (or), alternatively, in maps which map points in $\Omega$ to
points in $\Omega.$ Notice, that under such conditions the Lie algebra
$VectS^{1}$can \textbf{always} be constructed since its construction requires
only existence of some \textbf{open} interval around any point z$\in\Omega.$
But, by definition, the set $\Omega$ \textbf{is} open.

Let us discuss now the issue of central extension of $VectS^{1}.$ The need to
introduce the central extension of the Lie algebra $VectS^{1}$ is by no means
intrinsic just for this group. Already Schur developed general method of
constructing projective representations of finite groups about a hundred years
ago. The extension of his method to Lie groups is relatively straightforward
and is wonderfully presented in the book by Hamermesh$[98]$. The comprehensive
up to date summary of results in this direction could be found in the
encyclopedic work, Ref[99]. It is not our purpose to provide a review of these
results. We would like here to explain the physical motivations leading to the
projective representations of Lie groups since the central extension is
directly related to construction of these projective representations.

As is well known, there are actually two different ways to solve quantum
mechanical problems. The first one comes from mathematics of solving of 2nd
order ordinary differential equations while the second one comes from the
algebraic (group-theoretic) approach to the same problem. The projective
representations are naturally associated with the second approach. In
particular, let g$_{1}$ and g$_{2}$ be two elements of some Lie group G. One
can think of unitary representations associated with group G. That is one can
try to find a unitary operator $U(g)$, $g\in G,$ such that
\begin{equation}
U(g_{1})U(g_{2})=U(g_{1}\circ g_{2})\text{ .} \tag{7.18}%
\end{equation}
Such representation of the group G is called \textbf{vector} representation(by
analogy with finite dimensional space where the role of $U$ is being played by
finite matrices acting on vectors). In quantum mechanics, as is well known,
the wave function is determined with accuracy up to a phase factor. This means
that, along with Eq.(7.18), one can think of alternative way of writing the
composition law, e.g.
\begin{equation}
U(g_{1})U(g_{2})=\omega(g_{1},g_{2})U(g_{1}\circ g_{2})\text{ .} \tag{7.19}%
\end{equation}
Surely, one should require $\left|  \omega(g_{1},g_{2})\right|  =1.$ This then
allows us to write the factor $\omega(g_{1},g_{2})$ as
\begin{equation}
\omega(g_{1},g_{2})=\exp\{i\xi(g_{1},g_{2})\} \tag{7.20}%
\end{equation}
The phase factor $\xi(g_{1},g_{2})$ is associated with the topology of the
underlying group space. Finally, in our case of $DiffS^{1}$ the action of the
operator $U(g)$ on the vector $f(z)$ is given by Eq.s(7.11)-(7.13) so that the
composition law, Eq.(7.19), along with definition, Eq.(7.20), allows us to
obtain in a rather standard way$[100]$ the centrally extended Lie algebra
$VectS^{1}$ which is known as Virasoro algebra and it is given by
\begin{equation}
\left[  \hat{d}_{m},\hat{d}_{n}^{{}}\right]  =(m-n)\hat{d}_{m-n}^{{}}+\hat
{c}a(m,n) \tag{7.21}%
\end{equation}
where $\hat{c}$ is some number (related to the central charge) and the
two-cocycle $a(m,n)$ is related to $\xi(g_{1},g_{2})$ and can be easily
obtained explicitly by using the Jacoby identity and the commutation relations
given by Eq.(7.21). The final result can be written in the form suggested by
Kac$[56]$%
\begin{equation}
\left[  \hat{d}_{m},\hat{d}_{n}^{{}}\right]  =(m-n)\hat{d}_{m-n}^{{}}%
+\delta_{m,n}\frac{(m^{3}-n)}{12}c \tag{7.22}%
\end{equation}
with c being the central charge. For the developments presented below in this
paper it is very important to recognize the \textbf{physical reason} for the
emergence of the two-cocycle a(m,n). Nag and Verjovsky$[48]$ had demonstrated
that it is related to the quasisymmetric deformations of the projective
structures on $S^{1}$by diffeomorphisms. These structures were fully
classified in Ref[101]. Basically, they are associated with group of
M$\ddot{o}$bius transformatoions PSL(2,R)
\begin{equation}
x=\frac{ax+b}{cx+d} \tag{7.23}%
\end{equation}
on the \textbf{real} line. Study of the deformations of the projective
structure on the line which was initiated by Ahlfors and Beurling$[94]$ was
considerably developed by Carleson$[95]$ and Agard and Kelingos$[96]$ and
culminated in the work of Nag and Verjovsky$[48]$ . To make our presentation
self-contained, we would like now to summarize their results from the point of
view of ideas presented in this section. This summary is needed whenever one
is contemplating about the extension of the existing 2 dimensional results
related to CFT to higher dimensions (to be discussed in some detail in the
next section).

Consider a quotient T(1)=QS/PSL(2,R) that is the space of ''true''
quasisymmetric deformations which fix 3 points, e.g.say, 1,-1 and -i on
$S^{1}$, then T(1) is associated with \textbf{universal} Teichm\"{u}ller space
in a sense of Bers$[102].$ The space $M=DiffS^{1}/PSL(2,R)$ is embeddable
inside of T(1). The space $M$ can be equipped with the complex structure so
that it becomes infinite dimensional K\"{a}hler manifold. For the vectors
$v=\sum\limits_{m}v_{m}\hat{d}_{m}$ and $w=\sum\limits_{m}w_{m}\hat{d}_{m}$
\textbf{tangent} to $M$ at some point chosen as the origin one can construct
the K\"{a}hler metric g($v,w).$ The most spectacular result of Ref [48] lies
in the proof of the fact that the K\"{a}hler 2-form
\begin{equation}
\omega(v,w)=g(v,\tilde{J}w), \tag{7.24}%
\end{equation}
where $\omega(v,w)=\sum\limits_{n,m}v_{n}w_{m}a(m,n),$ with $a(m,n)$ being the
same as in Eq.(7.21), (7.22), and $\tilde{J}w$ being defined through equation
\begin{equation}
\tilde{J}w=\sum\limits_{m}(-i)sign(m)w_{m}\hat{d}_{m}, \tag{7.25}%
\end{equation}
coincides with the Weil-Petersson metric,
\begin{equation}
g(v,w)=\text{W-P}(v,w) \tag{7.26}%
\end{equation}
where W-P$(v,w)$ is the Weil-Petersson (W-P) metric on T(1).The Weil-Petersson
metric on Teicm\"{u}ller space is discussed in sufficient detail in Ref[45].If
$\mu(z)\frac{d\bar{z}}{dz}$ is the Beltrami differential,e.g. see Eq.(7.4),and
$\varphi([\nu])(z)dz^{2}$ is the quadratic differential (e.g. see Ref[37] for
an elementary discussion of quadratic differentials) then, the W-P inner
product is defined by the following formula
\begin{align}
&  <\mu,\varphi\lbrack\nu]>=\text{W-P}(\mu,\nu)\tag{7.27}\\
\text{ \quad\quad\quad\quad\quad}  &  =\text{{}}\iint\limits_{\Delta/F}%
\times\iint\limits_{\Delta}\frac{\mu(z)\bar{\nu}(\zeta)}{(1-z\bar{\zeta})^{4}%
}d\xi d\eta dxdy\nonumber
\end{align}
with $\nu\rightarrow\varphi([\nu])(z)$ being given by
\begin{equation}
\varphi([\nu])(z)=\iint\limits_{\Delta}\frac{\bar{\nu}(\zeta)}{(1-z\bar{\zeta
})^{4}}d\xi d\eta\tag{7.28}%
\end{equation}
, here$z=x+iy$ , $\zeta=\xi+i\eta$ and $\Delta$ is the unit disk with F
beingsome Fuchsian group thus making the quotient $\Delta/F$ a Riemann
surface. In the present case F$\equiv1$ as it will be explained shortly. In
view of this, one should not worry about F.

\textbf{Remark} \textbf{7.17}.a) The Kalerity of W-P metric expressed by
Eq.(7.24) had actually been proven by Ahlfors$[49]$ in 1961. b) In the same
paper by Ahlfors, Eq.(7.28) has been derived which differs in sign and
numerical prefactor from Eq.(7.28). This, fortunately, plays no role in the
final results obtained in Ref.[48].

\textbf{Remark 7.18} Since Eq.(7.28) plays the central role in the rest of
calculations presented below,we would like to provide some additional
information related to this equation(\textbf{not }contained in Ref[48]) in
order to help physically educated reader to appreciate its significance . To
this purpose let $\mu_{f}$ in Eq.(7.4) be written as $\mu(t)(z)=t\nu(z).$
Then, it can be shown [44] that for solution $f^{\mu}$ of the Beltrami
Eq.(7.4) the following limiting result holds
\begin{align}
\dot{v}[\mu](z)  &  =\lim_{t\rightarrow0}\frac{f^{\mu}-z}{t}\tag{7.29}\\
&  =-\frac{1}{\pi}\iint\limits_{H^{2}}\nu(z)\frac{z(z-1)}{\zeta(\zeta
-1)(\zeta-z)}d\xi d\eta\text{ .}\nonumber
\end{align}
With help of Eq.(7.29) we obtain,
\begin{equation}
f^{\mu}(z)=z+\dot{v}[\mu](z)t+o(t)\text{ , }t\rightarrow0, \tag{7.30}%
\end{equation}
to be compared with Eq.(7.12). From this comparison it follows, that the
quasisymmetric vector field $\mathbf{v}$ on $S^{1}$ can be defined as
\begin{equation}
v=\dot{v}[\mu](z)\frac{d}{dz}. \tag{7.31}%
\end{equation}
In addition, using Eq.(7.30), we obtain,
\begin{equation}
\frac{d}{dz}f^{\mu}(z)=1+t\frac{d}{dz}\dot{v}[\mu](z)\equiv1+t\dot{v}%
[\mu]^{\prime} \tag{7.32}%
\end{equation}
and also,
\begin{equation}
\frac{d^{2}}{dz^{2}}f^{\mu}(z)=t\dot{v}[\mu]^{\prime\prime} \tag{7.33}%
\end{equation}
and
\begin{equation}
\frac{d^{3}}{dz^{3}}f^{\mu}(z)=t\dot{v}[\mu]^{\prime\prime\prime}\text{ .}
\tag{7.34}%
\end{equation}
The Schwarzian derivative of $\{f^{\mu},z\}$ defined by
\begin{equation}
\varphi_{t}[\nu\}(z)=\{f^{\mu},z\}=\frac{f^{\prime\prime\prime}(z)}{f^{\prime
}(z)}-\frac{3}{2}\left(  \frac{f^{\prime\prime}(z)}{f^{\prime}(z)}\right)
^{2} \tag{7.35}%
\end{equation}
can now be constructed so that in the limit $t\rightarrow0$ using
Eqs(7.32)-(7.35) we obtain,
\begin{equation}
\varphi_{t}[\nu](z)=t\dot{v}[\mu]^{\prime\prime\prime}+o(t)\text{ .}
\tag{7.36}%
\end{equation}
Let now $\mu(t)=\mu+t\nu\lbrack z],$ then we can construct
\begin{align}
\varphi\lbrack z](z)  &  =\frac{\varphi_{t}[\nu](z)-\varphi_{0}[\nu](z)}%
{t}\tag{7.37}\\
&  =-\frac{12}{\pi}\iint\limits_{H^{2}}\frac{\bar{\nu}(z)}{(\bar{\zeta}%
-z)^{4}}d\xi d\eta\nonumber
\end{align}
where the use was made of Eq.(7.29) in order to perforn z-differentiation in
eq.(7.36) explicitly. Obtained result is documented on p.138 of Ahlfors book,
Ref[92], and should be compared against Eq.(7.28) upon conversion from H$^{2}$
plane to the disc $\Delta$. Since it is well known$[44]$ that the Schwarzian
derivative acts like a quadratic differential under the transformations which
belong to the Fuchsian group F, we conclude that, indeed, up to unimportant
constant (which may differ from $-\frac{12}{\pi}$ when the transformation from
H$^{2}$ to $\Delta$ is made ) Eqs.(7.28) and (7.37) are equivalent.

Next, by combining Eqs.(7.21),(7.24),(7.25) it can be shown, that
\begin{equation}
g(v,w)=-2i\hat{c}\operatorname{Re}\sum\limits_{m=2}^{\infty}\bar{v}_{m}%
w_{m}(m^{3}-m). \tag{7.38}%
\end{equation}
In addition, it is possible to show that the Fourier coefficients of $\dot
{v}[\mu]$ (and, analogously, $\dot{w}[\mu])$ defined by Eq.(7.31),are given
by
\begin{equation}
v_{k}=\frac{i}{\pi}\iint\limits_{\Delta}\bar{\mu}(z)z^{k-2}dxdy, \tag{7.39a}%
\end{equation}
and
\begin{equation}
w_{k}=\frac{i}{\pi}\iint\limits_{\Delta}\nu(z)z^{k-2}dxdy\text{ , k}\geq2.
\tag{7.39b}%
\end{equation}
Using these results in Eq.(7.38), we obtain,
\begin{equation}
\sum\limits_{m=2}^{\infty}\bar{v}_{m}w_{m}(m^{3}-m)=-\frac{1}{\pi^{2}}%
\iint\limits_{\Delta}\times\iint\limits_{\Delta}\mu(z)\bar{\nu}(\zeta
)(\sum\limits_{m=2}^{\infty}z^{m-2}\bar{\zeta}^{m-2}(m^{2}-m))d\xi d\eta dxdy
\tag{7.40}%
\end{equation}
Using summation formula
\begin{equation}
\sum\limits_{m=2}^{\infty}(m^{3}-m)x^{m-2}=\frac{-1}{6(1-x)^{4}}\text{
,}\left|  x\right|  <1 \tag{7.41}%
\end{equation}
in Eq(7.40) we obtain,
\begin{equation}
g(v,w)=-\frac{i\hat{c}}{3\pi^{2}}\text{W-P}(\mu,\nu) \tag{7.42}%
\end{equation}
with W-P$(\mu,\nu)$ being defined by Eq.(7.27) were now we have to put F=1.
Surely, $\hat{c}$ can be replaced by $ib$ and we can adjust $b$ in such a way
that $\frac{b}{2\pi^{2}}=\frac{c}{12}$ in accord with Eq.(7.22).

Thus, we have demonstrated, following Nag and Verjovsky$[48]$, that
\textbf{the central charge of the} \textbf{Virasoro algebra is directly
associated with the quasisymmetric deformations of} $\Delta$ (or H$^{2}).$In
view of this fact, it becomes possible to consider extensions of the existing
formalism to higher dimensions. This is the subject of the next section.

\textbf{Remark 7.18.} Since the Virasoro algebra, Eq.(7.22), with
\textbf{fixed} central charge provides solution of a particular CFT at
criticality, to crossover from one universality class (given by fixed central
charge) to another (given by different value of the central charge)
Zamolodchikov [103] had developed theory (known in physics literature as
c-theorem) which describes the dynamics of crossover between different values
of central charge. It would be very interesting to explain his results by
developing ideas of Nag and Verjovsky[48].

\mathstrut

{\Large 8. Beyond two dimensions}

\mathstrut

In 1968 Mostow$[104]$ proved very important theorem which is known as Mostow
rigidity theorem. It can be formulated as follows.

\textbf{Theorem} \textbf{8.1}.(Mostow)\textit{Let N=H}$^{d+1}/\Gamma
$\textit{\ be a complete hyperbolic manifold , d}$\geq2,$\textit{\ and let
N}$^{^{\prime}}=$\textit{H}$^{d+1}/\Gamma^{\prime}$\textit{\ be some other
hyperbolic manifold, then if there is a quasi-isometric homeomorphism f:
N}$\rightarrow$\textit{N}$^{^{\prime}}$\textit{, then f is homotopic to an
isometry N}$\rightarrow$\textit{N}$^{\prime}$\textit{\ if and only if both
M\"{o}bius groups }$\Gamma$\textit{\ and }$\Gamma^{\prime}$\textit{\ are of
the first kind (i.e. }$\Omega=S_{\infty}^{d}-\Lambda=0$\textit{,e.g. see
section 5).}

\textbf{Remark 8.2}.This result could be easily understood in view of Eq.s
(7.7) and (7.8). For an additional illustration of the existing possibilities
one is encouraged to look at the paper by Donaldson and Sullivan$[105]$ who
established that some closed four-manifolds have infinitely many distinct
quasiconformal structures while others do not admit the quasiconformal
structure at all.

\textbf{Remark} \textbf{8.3}.Mostow rigidity theorem can be viewed as an
extension and ramification of much earlier theorem by Liouville (originally
proven in 1850)$[106]$ which can be stated as follows.

\textbf{Theorem 8.4}.(Liouville) Let $U$ be some open subset of R$^{d}%
\cup\{\infty\}\equiv$\^{R}$^{d}$ and let f :$U\rightarrow$\^{R}$^{d}$ be a
conformal map, then f is just a M$\ddot{o}$bius transformation for d$\geq3.$

It is because of this theorem, known in physics literature$[9],$ there is a
widespread belief that results of two dimensional CFT\textbf{\ cannot} be
extended to higher dimensions.

\textbf{Remark} \textbf{8.5.} In order to study d dimensional systems at
criticality (d$\geq2)$ one should look for the M$\ddot{o}$bius groups of the
\textbf{second} kind . Then, the question arises immediatly : is there an
analogue of physically funadamentally important Canary-Taylor theorems
(Theorems 7.3. and 7.5) in higher dimensions? We are unaware of a
comprehensive answer to this question. However, we would like to mention the
''tour de force'' papers by Gromov, Lawson and Thurston$[107]$ and also by
Kuiper$[108]$ from which it follows that, at least for groups of isometries of
H$^{4}$ considered in these references, the limit set\textbf{\ is} a circle
$S^{1}$(actually, nowhere differentiable Julia-like set).

In view of the above lack of Canary-Taylor theorems in higher dimensions, we
would like to discuss now different methods of study of the limit sets (and
their complements) of M$\ddot{o}$bius groups in dimensions higher than 3. To
this purpose, using Eqs (5.2)-(5.5) and following McMullen$[27]$(and
Thurston$[42],$chapter 11$),$ we define the map :
\[
av:\text{ }C^{\infty}(S_{\infty}^{d},R)\rightarrow C_{{}}^{\infty}%
(\text{H}^{d+1},R)
\]
or
\begin{equation}
\mathcal{F(}0)\equiv av(f)(0)=\frac{1}{\omega_{d}}\int\limits_{S^{d}}%
d\omega(x)f(x) \tag{8.1}%
\end{equation}
i.e. the map $av$(f) is the average of f over $S_{\infty}^{d}.$ Using
Eqs.(5.2),(5.5), we obtain:
\begin{align}
\mathcal{F(}y)  &  =\mathcal{F(\gamma}0)=\frac{1}{\omega_{d}}\int
\limits_{S^{d}}d\omega(x)f(\gamma x)\tag{8.2}\\
&  =\frac{1}{\omega_{d}}\int\limits_{S^{d}}d\omega(x)f(T_{y}^{-1}x)\nonumber\\
&  =\frac{1}{\omega_{d}}\int\limits_{S^{d}}d\omega(x)\left|  T_{y}^{\prime
}(x)\right|  ^{d}f(x)\nonumber\\
&  =\frac{1}{\omega_{d}}\int\limits_{S^{d}}d\omega(x)\left(  \frac{1-\left|
y\right|  ^{2}}{\left|  x-y\right|  ^{2}}\right)  ^{d}f(x)\nonumber\\
&  =av(\gamma f)(0)\nonumber
\end{align}
here y$\in$H$^{d+1}$, T$^{-1}0=y.$

Using Eq.s (2.14),(3.7),(3.8),(3.15) and (5.2) we conclude that
\begin{equation}
\Delta_{h}av(\gamma f)(0)=0. \tag{8.3}%
\end{equation}
That is the avarage $av(f)$ is a harmonic function in hyperbolic metric. It is
clear, that to restore the harmonic function $\mathcal{F}(x)$ in H$^{d+1}$ it
is sufficient to know the function $f(x)$ at the boundary of hyperbolic space,
i.e. on $S_{\infty}^{d}($ recall the holography principle discussed in section 1)

Let now $\frak{v(}x)$ be some vector field, $\frak{v(}x)\in S_{\infty}^{d}%
.$Then, as before, one can extend it to the bulk of hyperbolic space by using
the prescription:
\begin{equation}
av(\frak{v})(0)=\frac{1}{\omega_{d}}\int\limits_{S^{d}}d\omega(x)\frak{v}(x)
\tag{8.4}%
\end{equation}
In the case of functions, $av(f)$ \textbf{by design} provides a continuous
function on $S_{\infty}^{d}\cup$H$^{d+1}.$ This is not true for vectors (or
tensors in general). In the case of vectors, one defines the
\textbf{extension} operator $ex(f)$ via the following prescription:
\begin{equation}
ex(f)=av(f)\text{ \ for scalar fields(functions),} \tag{8.5}%
\end{equation}%
\begin{equation}
ex(\frak{v})=\frac{d+1}{2d}av(\frak{v})\text{ \quad for vector fields, etc.}
\tag{8.6}%
\end{equation}
Being armed with these results we are ready to extend the results of previous
section to higher dimensions. To this purpose, we need to reanalyze Eq.(7.29)
first. It is equation for the \textbf{vector} field which is created by
deformation $\nu(\zeta).$ It can be shown, e.g.see pages 196-197 of Ref[44],
that
\begin{equation}
\frac{\partial}{\partial\bar{z}}\dot{v}[\mu](z)=\nu(z) \tag{8.7}%
\end{equation}
that is when $\nu(z)=0,\dot{v}[\mu](z)$ is just a holomorphic function which
obeys the Cauchy-Riemann equations. Ahlfors had demonstrated $[25]$that, there
is an analogue of Eq.(8.7) in higher dimensions. Let f$_{i}(x)=\dot{v}_{i}%
[\mu](x),$ $x\in S_{\infty}^{d}\cup$H$^{d+1}$then, the higher dimensional
analogue of Eq.(8.7) is given by
\begin{equation}
\left(  Sf\right)  _{ij}=\frac{1}{2}\left(  \frac{\partial f_{i}}{\partial
x_{j}}+\frac{\partial fj}{\partial x_{i}}\right)  -\frac{\delta_{ij}}{d+1}%
\sum\limits_{k=1}^{d+1}\frac{\partial f_{k}}{\partial x_{k}}=\Xi_{ij}(x).
\tag{8.8}%
\end{equation}
It can be shown, that Eq.(8.8) is reduced to Eq.(8.7) in two dimensions. In a
special case $\Xi_{ij}(x)=0$ one obtains solution of Eq.(8.8) in the form
\begin{equation}
f_{i}^{0}=a_{i}+\sum\limits_{j}A_{ij}x_{j}+b_{i}\mathbf{x}^{2}%
-2(\mathbf{b\cdot x})x_{i} \tag{8.9}%
\end{equation}
where $\mathbf{a}$ and \textbf{\ }$\mathbf{b}$ are some constant vectors and
$\mathbf{A}$ is a constant matrix which is the sum of skew-symmetric and
diagonal (with the same elements along the diagonal)matrices. Apart from the
matrix term in Eq.(8.9), the above result is identical with that known in
physics literature, (e.g.see Ref.[9], Eq.(4.14)). By analogy with Eq.(7.31) in
two dimensions (taking into account the behaviour at infinity$[27]$) one
obtains for the vector field
\begin{equation}
v(z)=(a+bz+cz^{2})\frac{\partial}{\partial z} \tag{8.10}%
\end{equation}
which clearly obeys $VectS^{1}$ Lie algebra, Eq.(7.15), as expected.The
central extension of this algebra given by Eq.s(7.21),(7.22) is \textbf{not}
affected by this field since for indices 1,0,-1 one has $a(m,n)=0.$ This is
also in complete accord with Eq.(7.38). This observation has very important
consequences. In particular, if one would like to obtain solution to Eq.(8.8)
for $\Xi\neq0,$then, obviously, the general solution $f_{i}$ is going to be
given by
\begin{equation}
f_{i}=f_{i}^{0}+\varphi_{i}\text{ .} \tag{8.11}%
\end{equation}
Hence, physically interesting nontrivial solutions of Eq.(8.8) are given by
$\varphi_{i}=f_{i}-f_{i}^{0}.$ This obseravtion can be broadly generalized
from the point of view of cohomology theory to be discussed briefly below. In
the meantime, one is faced with the problem of finding solutions to Eq.(8.8)
for $\Xi\neq0$ . Ahlfors$[25]$ had found a very ingenious way of doing this.
To this purpose, he had introduced the operator $S^{\ast}$ adjoint to $S$.
Without going into details of its explicit form which could be found in his
work, the main point of having such an operator lies in selecting such
$\Xi(s)$ for which
\begin{equation}
S^{\ast}\Xi=0\text{ .} \tag{8.12}%
\end{equation}
Then, by analogy with results of sections 2 and 3, one obtains the following
Dirichlet-type problem of finding the solutions of the Laplace-like equation
in B$^{d+1}:$%
\begin{equation}
\rho^{-d-3}S^{\ast}\rho^{d+1}S\mathbf{v}=0\text{ , }\rho=\frac{1}{1-x^{2}%
}\text{\quad,} \tag{8.13}%
\end{equation}
supplemented with the boundary condition
\begin{equation}
\mathbf{v}\mid_{S_{\infty}^{d}}=\mathbf{f}\text{ , }x^{2}=1\text{ at
}S_{\infty}^{d}. \tag{8.14}%
\end{equation}
To solve this equation, one has to assign the vector fields at the boundary. A
complete solution which takes into account Eqs.(8.5),(8.6) was obtained by
Reiman$[109].$ An alternative derivation which uses the theory of pseudo-
Anosov homeomorphisms (which we had discussed in connection with dynamics of
2+1 gravity and textures in liquid crystals[36],[37]) was recently obtained by
Kapovich$[110].$ He proved the following

\textbf{Theorem 8.6}.(Kapovich) \textit{Suppose that \textbf{v} is a smooth
automorphic k-quasiconformal vector field on the open unit ball B}$^{d+1}%
$\textit{\ in R}$^{d+1},d\geq2.$\textit{\ Then \textbf{v} admits a continuous
tangential extension }\textbf{v}$_{\infty}$\textit{\ to }$S_{\infty}^{d}.$
\textit{The vector field }\textbf{v}$_{\infty}$\textit{\ is again a
k-quasiconformal vector field on the sphere }$S_{\infty}^{d}.$

\textbf{Remark 8.7.}Recent attempts [1],[111] to extend CFT theories to higher
dimensions for technical reasons are limited to even dimensionalities, e.g. 2,
4 and 6. The results of Reiman and Kapovich can be used for any d$\geq2.$This
fact is consistent with latest results of Bakalov et al [57].

To have some appreciation of these more general results, our experience with
two dimensional case discussed in section 7 is helpful. It is also useful for
development of cohomological methods[112] of study of deformations of Kleinian
(and, in general, M$\ddot{o}$bius ) groups.We shall follow mainly the ideas of
Refs[44]and[113] since, in our opinion, they are the most helpful for
undrstanding of more sophisticated treatments$[112],[114]$ not limited to
dimension two.

The starting point is Eq.(8.7). If $\dot{v}[\mu](z)\equiv F(z)$ is a vector
field, then, naturally, we have to require
\begin{equation}
F(\gamma\circ z)=\gamma^{\prime}F(z), \tag{8.15}%
\end{equation}
where $\gamma^{\prime}$ was defined after Eq.(5.4). Following Ref[45], let us
call $F(z)$ a ''potential'' for $\nu$. It is clear, that Eq.(8.7) must be
consistent with Eq.(8.15). This imposes some restrictions on the potential $F$
that is we have to demand that the combination $F(\gamma\circ z)-\gamma
^{\prime}F(z)$ vanishes for any $\gamma\in\Gamma.$ Define now the function
$\chi_{F}(\gamma)=\left(  F(\gamma\circ z)/\gamma^{\prime}\right)  -F$. Taking
into account Eqs(7.29),(7.31),(8.7),(8.9) and (8.10), we conclude, that vector
$\chi_{F}(\gamma)$ should be proportional to that given in Eq.(8.10). At the
same time, it should satisfy the one-cocycle condition
\begin{equation}
\chi_{F}(\gamma_{1}\circ\gamma_{2})=\left(  \gamma_{2}\right)  _{\ast}%
(\chi_{F}(\gamma_{1}))+\chi_{F}(\gamma_{2})\text{ , }\gamma_{1},\gamma_{2}%
\in\Gamma\tag{8.16}%
\end{equation}
with
\begin{equation}
\gamma_{\ast}(P)=\frac{P\circ\gamma}{\gamma^{\prime}}\text{ .} \tag{8.17}%
\end{equation}
Indeed, since we have
\[
\chi_{F}(\gamma_{1})=\frac{F\circ\gamma_{1}}{\gamma_{1}^{\prime}}-F
\]
and
\[
\chi_{F}(\gamma_{2})=\frac{F\circ\gamma_{2}}{\gamma_{2}^{\prime}}-F,
\]
we expect that
\[
\chi_{F}(\gamma_{1}\circ\gamma_{2})=\frac{F\circ(\gamma_{1}\circ\gamma_{2}%
)}{\left(  \gamma_{1}^{{}}\circ\gamma_{2}\right)  ^{^{\prime}}}-F.
\]
Use of these results in Eq.(8.16) produces the result which is well known in
the theory of dynamical systems$[97]:$%
\begin{equation}
\left(  \gamma_{1}\circ\gamma_{2}\right)  ^{\prime}=\gamma_{1}^{^{\prime}%
}\cdot\gamma_{2}^{^{\prime}}\text{ .} \tag{8.18}%
\end{equation}
Let $W$ be another potential for $\nu,$ then $P=W-F$ is again proportional to
the vector field given by Eq.(8.10). Thus, $\chi_{W}-\chi_{F}=\delta(P)$ where
$\delta(P)(\gamma)=\gamma_{\ast}(P)-P.$ Recall now that, according Eq.(7.29),
we had defined $\mu(t)(z)=t\nu(z).$ Therefore Eq.(7.9) can be rewritten as
$\gamma^{t}=f^{t}\circ\hat{\gamma}\circ(f^{t})^{-1}$ so that $\frac{d}%
{dt}\gamma^{t}\mid_{t=0}=\dot{\gamma}$. By combining this result with
Eq.(7.29) we obtain,
\begin{equation}
\frac{df^{t}}{dt}\mid_{t=0}=F(z)\equiv\dot{f}[\mu], \tag{8.19}%
\end{equation}
and also,obviously,
\begin{equation}
f^{t}\circ\hat{\gamma}=\gamma^{t}\circ f^{t}. \tag{8.20}%
\end{equation}
Differentiating the last equation and, again, taking into account Eq.(7.29) we
obtain,
\begin{equation}
\dot{f}[\mu]\circ\gamma=\dot{\gamma}+\dot{f}[\mu]\cdot\gamma^{\prime}.
\tag{8.21}%
\end{equation}
This leads us to
\begin{equation}
\chi_{\dot{f}[\mu]}=\frac{\dot{f}[\mu]\circ\gamma}{\gamma^{\prime}}-\dot
{f}[\mu]=\frac{\dot{\gamma}}{\gamma}. \tag{8.22}%
\end{equation}
In view of Eq.(8.15) we observe that the obtained result is nontrivial.
Accordingly, if $\chi_{\dot{f}[\mu]}$ is the vector space $Z^{1}$ of cocycles
and $\delta(P)$ is the vector space $B^{1}$ of coboundaries, then the
quotient
\begin{equation}
H^{1}=Z^{1}/B^{1} \tag{8.23}%
\end{equation}
defines the first Eichler cohomology group of $\Gamma$, that is the group of
\textbf{nontrivial} deformations. With some efforts$[113]$ it is possible to
construct the second and higher Eihler cohomology groups. Although the above
analysis seems quite natural, the higher dimensional generalizations of such
cohomological arguments so far had been based on the cohomology theory
developed by Eilenberg and MacLane $[115],$e.g.see Ref[112], which is
conceptually similar but technically a bit different from the Eichler
theory$[113].$ The reasons for such limitations of Eichler's approach are
clear : all arguments use two dimensional complex analysis. In our opinion,
Eilenberg-MacLane approach is more formal and, hence, allows much lesser use
of physical intuition. The famous Gelfand-Fuks two- cocycle obtained with help
of Eilenberg-MacLane cohomology theory (also in a rather formal way) for the
Lie algebra of the vector fields is known to produce the central extension of
the $VectS^{1}$ Lie algebra $[58],$e.g. see eq.(7.15). Recently, the authors
of Ref.[57] had sucseeded in consistently extending the cohomological results
\ of Gelfand and Fuks to higher dimensions (although \ some work is still in
progress).It remains a challenging problem to connect these results with the
cohomological results of Johnson and Millson[112] and Kourouniotis [116] which
take explicitly into account deformations of hyperbolic groups. In
anticipation of more rigorous mathematical results, we would like to present
now some more intuitive physical -type arguments which enable us to provide
some answers to these problems.

First, we have to think about the higher dimensional analogue of the Lie
algebra for the group PSL(2,C). In two dimensions it forms a closed subalgebra
within the Virasoro algebra. For concretness, let us think about description
of 3 dimensional conformal models, i.e.d+1=4 . As it was shown by
Cartan$[50],$ the Lie algebra of conformal transformations of $\mathbf{R}%
^{d+1}$ is isomorphic to the Lie algebra of the group O(d+1,1). For our
purposes we need actually only the component connected to identity S$O_{0}%
$(d+1,1) of O(d+1,1). As it was shown recently by Scannell$[60]$, this group
is \textbf{simultaneously isomorphic} to the group $Isom^{+}($H$^{d+1})$ which
is group of orientation -preserving isometries of H$^{d+1},$ the group
$M\ddot{o}b^{+}(S^{d})$of orientation -preserving M$\ddot{o}$bius
transformations of $S_{\infty}^{d}$ and the group $Isom^{+}(S_{1}^{d+1})$ of
isometries of the de Sitter space ($S_{1}^{d+1}=\{\mathbf{v\in R}_{1}%
^{d+2}\mid<\mathbf{v},\mathbf{v}>=1\}$ with \textbf{R}$_{1}^{d+2}$ being the
space \textbf{R}$^{d+2}$ equipped with the signature (d+1,1)). We shall use
the last option for reasons which will become obvious momentarily.
Incidentally, for d=2 we have to deal with the group SO(3,1) which is just the
Lorentz group isomorphic to PSL(2,C) as discussed in great detail in Ref[[51].
It is very striking that the representations of the Lie group SO(4,1) and, in
particular, its connected component, describe the spectrum of the hydrogen
atom$[55].$ This fact is helpful for treatment of 3d conformal models. From
the detailed analysis of the de Sitter group performed in Refs[52],[53] it
follows, that the Lie algebra of the group SO(4,1) is isomorphic to the direct
product of two Lie algebras of the group SO(3),i.e. $so(4,1)$=$so(3)\otimes
so(3).$ But it is well known that Lie algebra $so(3)$ can be mapped onto
PSL(2,C) (it is intuitively clear since via stereographic projection the
sphere S$^{2},$ on which $so(3)$ acts, is being mapped onto the extended
complex plane(on which PSL(2,C) acts) and, indeed, the commutation relations
given by Eq.s (4.3) of Ref[52], up to a trivial rescaling, coinside exactly
with those given by Eq.(7.15). Since $VectS^{1}$ Lie algebra, Eq.(7.15),
admits central extension, we, thus, arrive at the direct product of two
Virasoro algebras which may have \textbf{different} central charges in
general. The task now is to find the highest weight representations for such
tensor product of two Virasoro algebras. This task makes sence to discuss only
if the limit set $\Lambda$ is union of two independent circles. In the light
of the results of Gromov, Lawson and Thurston [107] for some four-manifolds
the limit set is, still, just a circle $S^{1}.$ Balinskii and Novikov$[117]$
had proposed the multicomponent extension of the Virasoro algebra (e.g. see
Eq.(14) of Ref[117]). Their work considers the embedding of $S^{1}$into
n-$\dim$ensional smooth manifold M, i.e.f : $S^{1}\rightarrow M$ ,
f(x)=\{u$^{i}(x),1\leq i\leq n;x\in S^{1}\}.$ Accordingly, there is
\textbf{only one} central charge. The cohomological analysis of this embedding
is discussed in a recent survey by Mokhov[47]. Apparently, the results of
Bakalov et al [57] are different from that discussed by Mokhov. Full analysis
of the emerging possibilities is left for future work.

\mathstrut

\textbf{Acknowelegements}.During the course of this work we had contacted many
people.The author would like to apologize if unintentionally he had forgotten
some names.Dick Canary (U.of Michigan) had provided us with many of his
published and yet unpublished papers, Katsuhiko Matsutaki (Ochanomizu
U.,Japan) had kindly informed us about his book (together with M.Taniguchi) on
hyperbolic manifolds just published by Oxford U.Press, Kenneth Stephenson
(U.of Tennessee , Knoxville) had provided us with an unpublished lecture notes
by Chris Bishop (Stony Brook) on Kleinian groups,3-manifolds,etc, Mike Wolf
(Rice University) had connected us with Kevin Scannell (U.of St.Louis) whose
recently finished PhD thesis and current work on de Sitter and anti de Sitter
spaces had been helpful ,Albert Marden (U.of Minnesota ) had provided us with
(yet unpublished) fundamental paper on the monodromy groups of Schwarzian
equations on compact Riemann surfaces,Jay Jorgensen (U.of Oklahoma) had
provided us with yet unpublished works on spectral properties of 3-manifolds,
Lou Kauffman (U.of Illinois at Chicago) had been very attentive to the course
of this work and had asked many hard questions which eventually helped to
improve the presentation , Dave Crockett (Clemson U.) had provided some
assistance with graphics, finally, the author acknowledges an unlimited
support of his wife, Sophia, whose relentless trips into the math library of
the University of Chicago made this work actually possible. When this work was
already completed Prof.W. Thurston had kindly informed us about useful web
sourcehttp://www.maths.warwick.ac.uk where under the section ''Journals and
Monographs'' one can find \ additional up to date useful information related
to the subject matters of this paper.

\pagebreak 

\bigskip

\textbf{References}

[1] \ \ \ A.Popov, ''Holomorphic Chern-Simons-Witten theory:

\ \ \ \ \ \ \ \ from 2d to 4d conformal field theories'', hep-th/9806239.

[2] \ \ \ E.Fradkin and M.Palchik, Phys. Reports \textbf{300} (1998) 1.

[3] \ \ \ E.Witten,Comm. Math.Physics\ \textbf{121} (1989) 351.

[4] \ \ \ E.Witten, ''The central charge in 3 dimensions'', in \textit{Physics and}

\ \ \ \ \ \ \ \ \textit{Mathematics of Strings,}World Scientific,Singapore ,1990.

[5] \ \ \ G.More and N.Seiberg, Comm.Math.Physics \textbf{123} (1989)177.

[6] \ \ \ L.Funar, Comm. Math. Physics \textbf{171} (1995) 405.

[7] \ \ \ R.Bing and V.Klee, J.London Math.Soc.\textbf{39} (1964) 86.

[8] \ \ \ B.Maskit, \textit{Kleinian Groups}, Springer-Verlag, Berlin,1987.

[9] \ \ \ P. Di Francesco,P.Mathieu and D.Senechal ,\textit{Conformal Field}

\ \ \ \ \ \ \ \ \ \textit{Theory}, Springer-Verlag, Berlin ,1997.

[10] \ \ A.Beardon and B.Maskit, Acta Math. \textbf{132} (1974) 1.

[11] \ \ \ J.Maldacena ,''The large N limit of superconformal field theories

\ \ \ \ \ \ \ \ \ \ and supergravity'', hep-th/9711200.

[12] \ \ \ D.Friedman,S Mathur, A.Matusis and L.Rastelli, ''Correlation

\ \ \ \ \ \ \ \ \ \ functions in the CFT$_{d}$/AdS$_{d+1}$ correspondence'', hep-th/9804058.

[13] \ \ \ E.Witten, ''Anti de Sitter space and holography'', hep-th/9802150.

[14] \ \ \ I.Aref'eva and I.Volovich, Phys.Lett.\textbf{ B 433 (}1998)49..

[15] \ \ \ W.M\"{u}ck and K.Viswanathan, Phys.Rev.\textbf{D 58} (1998) 041901.

[16] \ \ \ W.M\"{u}ck and K.Viswanathan, Phys.Rev.\textbf{D 58} (1998) 106006.

[17] \ \ \ \ L.Suskind , J.Math.Phys.\textbf{36} (1995) 6377.

[18] \ \ \ \ G.'t Hooft, ''Dimensional reduction in quantum gravity''

\ \ \ \ \ \ \ \ \ \ , qr-qc/9310026.

[19] \ \ \ \ L.Suskind and E.Witten, ''The holographic bound in anti de-Sitter

\ \ \ \ \ \ \ \ \ \ \ space'', hep-th/9805114.

[20] \ \ \ \ C.Fefferman and C.Graham, ''Conformal invariants'', pp.92-116

\ \ \ \ \ \ \ \ \ \ in Asterisque (hors serie),1985.

[21] \ \ \ \ C.Graham and J.Lee,Adv.Math.\textbf{87} (1991) 186.

[22] \ \ \ \ A.Petrov, \textit{Einsein Spaces}, Pergamon, London, 1969.

[23] \ \ \ \ \ T.Willmore, \textit{Riemannian Geometry}, Clarendon Press, Oxford,1993.

[24] \ \ \ \ R.Kulkarni,''Conformal structures and M\"{o}bius structures'',

\ \ \ \ \ \ \ \ \ \ \ \ pp.1-39 in \textit{Conformal} \textit{Geometry},
F.Vieveg\&Sohn, Wiesbaden,1988.

[25] \ \ \ \ L.Ahlfors, \textit{M}$\ddot{o}$\textit{bius Transformations in
Several Dimensions ,}

\ \ \ \ \ \ \ \ \ \ \ \ Lecture Notes ,U.of Minnesota \ ,1981.

[26] \ \ \ \ S.Hawking and G.Ellis, \textit{The Large Scale Structure of
Space-Time, }

\ \ \ \ \ \ \ \ \ \ Cambridge U.Press, Cambridge,1973.

[27] \ \ \ \ C.McMullen , Renormalization and 3-manifolds Which Fiber

\ \ \ \ \ \ \ \ \ \ \ \ Over the Circle,\ Princeton U.Press, Princeton, 1996.

[28] \ \ \ \ J.Elstrodt, F.Grunevald and J.Mennicke ,\textit{Groups Acting on}

\ \ \ \ \ \ \ \ \ \ \textit{\ Hyperbolic Space,} Springer-Verlag , Berlin, 1998.

[29] \ \ \ \ J.Dozdziuk and J.Jorgenson, Memoirs of AMS \textbf{643} (1998).

[30] \ \ \ \ D.Sullivan, J.Differential Geometry \textbf{18} (1983) 723,

\ \ \ \ \ \ \ \ \ \ \ and also T.Lyons and D.Sullivan, J.Diff.Geometry
\textbf{19} (1984) 299.

[31] \ \ \ K.Matsutaki and M.Taniguchi, \textit{Hyperbolic Manifolds and Kleinian}

\ \ \ \ \ \ \ \ \ \ \ \textit{Groups}, Oxford U.Press, Oxford ,1998.

[32] \ \ \ \ P.Nicholls, \textit{The Ergodic Theory of Discrete Groups,}

\ \ \ \ \ \ \ \ \ \ \ Cambridge U.Press , Cambridge ,1989 .

[33] \ \ \ \ J.Morrow and K.Kodaira, \textit{Complex Manifolds}, Holt et all

\ \ \ \ \ \ \ \ \ \ \ Inc., New York ,1971.

[34] \ \ \ A.Presley and G.Segal, \textit{Loop Groups}, Oxford U.Press,

\ \ \ \ \ \ \ \ \ \ Oxford ,1986 .

[35] \ \ \ \ A.Belavin,A.Polyakov and A.Zamolodchikov,

\ \ \ \ \ \ \ \ \ \ \ Nucl. Phys. \textbf{B 241(}1984) 333.

[36] \ \ \ \ A.Kholodenko, ''Use of meanders and train tracks for

\ \ \ \ \ \ \ \ \ \ \ description of defects and textures in liquid crystals and

\ \ \ \ \ \ \ \ \ \ \ 2+1 gravity'' , hep-th/9901040 (JGP, in press).

[37] \ \ \ \ A.Kholodenko,''Use of quadratic differentials for description

\ \ \ \ \ \ \ \ \ \ \ of defects and textures in liquid crystals and 2+1
gravity'' ,

\ \ \ \ \ \ \ \ \ \ \ hep-th/9901047 (JGP,in press).

[38] \ \ \ \ J.Douglas ,Am.J.of Mathematics \textbf{61}(1939) 545.

[39] \ \ \ \ S.Axler, P.Bourdon and W.Ramey , \textit{Harmonic Function Theory}

\ \ \ \ \ \ \ \ \ \ \ ,Springer- Verlag, Berlin,1992 .

[40] \ \ \ \ S.Patterson,\ Acta Math.\textbf{136 }(1976\textbf{)} 241.

[41] \ \ \ \ \ D.Sullivan, I.H.E.S .Publ. \textbf{50 (}1979\textbf{)} 171.

[42] \ \ \ \ W.Thurston , \textit{Geometry and Topology of 3-Manifolds},

\ \ \ \ \ \ \ \ \ \ \ Princeton U.Lecture Notes, 1979. (http://www.msri.org/gt3m/)

[43] \ \ \ C.Bishop and P.Jones, Acta Math. \textbf{179} (1997)1.

[44] \ \ \ \ Y. Imayoshi and M.Taniguchi , \textit{An Introduction to
Teichm\"{u}ller}

\ \ \ \ \ \ \ \ \ \ \ \textit{Spaces}, Sringer-Verlag, Berlin ,1992 .

[45] \ \ \ \ L.Bers, Bull.London Math.Soc.\textbf{4 (}1972\textbf{)} 257.

[46] \ \ \ \ R.Canary and E.Taylor,\ Duke Math.Journal \textbf{73} (1994) 371.

[47] \ \ \ \ O.Mokhov, Russian Math. Surveys\textbf{ 53} (1998) 86 (in Russian).

[48] \ \ \ \ S.Nag and A. Verjovsky, Comm.in Math.Physics \textbf{130 (}1990) 123.

[49] \ \ \ \ L.Ahlfors, Ann. of Math. \textbf{74 }(1961)171 .

[50] \ \ \ \ E.Cartan, ''Les espaces $a^{\prime}$ connexion conforme'',

\ \ \ \ \ \ \ \ \ \ \ Ouevres Complete Vol.1, part III, pp.747-797 ,

\ \ \ \ \ \ \ \ \ \ \ Gauthier-Villars, Paris,1955.

[51] \ \ \ \ \ I.Gelfand, P.Milnos and Z.Shapiro, \textit{Representations of the}

\ \ \ \ \ \ \ \ \ \ \ \textit{Rotation}\ \textit{and Lorenz Groups }, Nauka,
Moscov ,1958 ( in Russian).

[52] \ \ \ \ \ L.Thomas, Ann.Math.\textbf{ 42} (1941)113.

[53] \ \ \ \ \ T.Newton, Ann.Math.\textbf{51} (1950)730 .

[54] \ \ \ \ \ E.Sudarshan, N.Mukuda and L.O'Raiifeartaigh,

\ \ \ \ \ \ \ \ \ \ \ Phys.Lett. \textbf{19} (1965) 322 .

[55] \ \ \ \ \ H.Bacry , Il Nuovo Cim. \textbf{16} (1966) 322 .

[56] \ \ \ \ \ V.Kac and A.Raina , \textit{Highest Weight Representations of Infinite}

\ \ \ \ \ \ \ \ \ \ \ \ \textit{Dimensional Lie Algebras },World Scientific, Singapore,1987.

[57] \ \ \ \ \ B.Bakalov,V.Kac and A.Voronov,Comm.

\ \ \ \ \ \ \ \ \ \ \ Math.Phys.\textbf{200}(1999)561.

[58] \ \ \ \ I. Gelfand and D.Fuks, Funct. Analysis and Appl.

\ \ \ \ \ \ \ \ \ \ \textbf{2 }(1968) 92 (in Russian).

[59] \ \ \ \ I.Aref'eva and I.Volovich ,''On large N conformal theories ,

\ \ \ \ \ \ \ \ \ \ field theories in anti de Sitter space and singletons'' ,

\ \ \ \ \ \ \ \ \ \ hep-th/9803028.

[60] \ \ \ \ K.Scannell, \textit{Flat Conformal Structures and Causality in de}

\ \ \ \ \ \ \ \ \ \ \textit{Sitter Manifolds}, PhD Thesis, U.of California
,Los Angeles,1996.

[61] \ \ \ A.Kholodenko, J.Math.Physics \textbf{37}(1996)1287.

[62] \ \ \ C. Itzykson and J-M Drouffe,\textit{ Statistical Field Theory} ,Vol.1,

\ \ \ \ \ \ \ \ \ \ Cambridge U.Press , Cambridge, 1989.

[63] \ \ \ A. Beardon, \textit{The Geometry of Discrete Groups}, Springer-

\ \ \ \ \ \ \ \ \ \ Verlag, Berlin, 1983 .

[64] \ \ \ A.Beardon in \textit{Ergodic Theory,Symbolic Dynamics and}

\ \ \ \ \ \ \ \ \ \ \textit{Hyperbolic Spaces}, \ T.Bedford, M.Keane and C.Series

\ \ \ \ \ \ \ \ \ \ Editors,Oxford U.Press,Oxford (1992).

[65] \ \ \ B.Mandelbrot , \textit{The Fractal Geometry of Nature} , W.Freeeman

\ \ \ \ \ \ \ \ \ \ Co.,\ New York ,1982.

[66] \ \ \ D.Sullivan, Acta Math.\textbf{153} (1984) 259.

[67] \ \ \ P.Tukia, Acta Math. \textbf{152} (1984)127.

[68] \ \ \ \ S.Helgason ,Groups and Geometric Analysis, Academic Press,

\ \ \ \ \ \ \ \ \ \ \ New York ,1984.

[69] \ \ \ \ A.Beardon, Proc.London.Math.Soc.\textbf{18} (1968) 461.

[70] \ \ \ \ A.Kholodenko and T.Vilgis, Phys.Reports \textbf{298} (1998) 251.

[71] \ \ \ \ B.Bowditch,\ J.of Functional Analysis \textbf{113} (1993) 245 .

[72] \ \ \ \ S.Patterson,Compositio Math. \textbf{31(}1975) 83;ibid
\textbf{32(}1976\textbf{)}71;

\ \ \ \ \ \ \ \ \ \ \ \ \ ibid \textbf{33}(1976) 227.

[73] \ \ \ \ D.Sullivan,\ J.of Differential Geom. \textbf{25} (1987)327.

[74] \ \ \ \ S.Patterson, Tohoku Math.Journ.\textbf{35} (1983) 357.

[75] \ \ \ \ E.Davies, B.Simon and M.Taylor, J.of Funct.

\ \ \ \ \ \ \ \ \ \ \ Analysis \textbf{78 }(1988)116.

[76] \ \ \ \ P.Lax and R.Phillips,Comm.on Pure and Appl.Math.\textbf{37}%
(1984)303 .

[77] \ \ \ \ C.Epstein, Memoirs of AMS \textbf{335} (1985).

[78] \ \ \ \ M.Burger and R.Canary ,\ J.Reine Angew. Math. \textbf{454} (1994) 37.

[79] \ \ \ \ R.Canary, Y.Minsky and E.Taylor ,''Spectral theory ,

\ \ \ \ \ \ \ \ \ \ \ Hausdorff dimension and the topology of hyperbolic

\ \ \ \ \ \ \ \ \ \ 3-manifolds'' , J.of Geometric Analysis (1999) (to be published)

[80] \ \ \ A.Grigoryan and M.Noguchi, Bull.London Math.Soc.\textbf{30}%
(1998)643 .

[81] \ \ \ W.Massey , \textit{Algebraic Topology: An Itroduction,}

\ \ \ \ \ \ \ \ \ \ Springer-Verlag, Berlin ,1967 .

[82] \ \ \ R.Benedetti and C.Petronio,\textit{ Lectures on Hyperbolic
Geometry},

\ \ \ \ \ \ \ \ \ \ Springer-Verlag, Berlin,1992.

[83] \ \ \ \ H.Farkas and I.Kra, \textit{Riemann Surfaces}, Springer-Verlag,

\ \ \ \ \ \ \ \ \ \ Berlin, 1992.

[84] \ \ \ \ L.Ahlfors, Am.J.of Math. \textbf{86 (}1964\textbf{)} 413.

[85] \ \ \ \ D.Sullivan, Acta Math. \textbf{147} (1981) 289.

[86] \ \ \ \ W.Abikoff, Proceedings AMS \textbf{97 (}1986\textbf{)} 593.

[87] \ \ \ \ S.Katok, \textit{Fuchsian Groups},U.of Chicago Press ,

\ \ \ \ \ \ \ \ \ \ \ Chicago,1992 .

[88] \ \ \ \ R.Bowen, I.H.E.S.Publ. \textbf{50} (1979)11.

[89] \ \ \ \ L.Bers, Proc.Symp.in Pure\ Math. \textbf{28} (1976) 559.

[90] \ \ \ \ W.Abikoff and B.Maskit, Am.J.Math.\textbf{ 99 (}1977\textbf{)} 687.

[91] \ \ \ \ J.Polchinski, \textit{String Theory} ,Vols I and II, Cambridge

\ \ \ \ \ \ \ \ \ \ \ U.Press,Cambridge ,1998.

[92] \ \ \ L.Ahlfors, \textit{Lectures on Quasiconformal Mappings,}

\ \ \ \ \ \ \ \ \ \ \ D.van Nostrand Co, New York,1966 .

[93] \ \ \ F.Gardiner and D.Sullivan, Am.J.Math.\textbf{114(}1992\textbf{)} 683.

[94] \ \ \ A.Beurling and L.Ahlfors, Acta Math.\textbf{96 }(1956)124.

[95] \ \ \ L.Carleson,\ J. d'Analyse Math.\textbf{19 (}1967\textbf{)}1.

[96] \ \ \ S.Agard and J.Keling, Comm.Math.Helv.\textbf{44} (1969) 446.

[97] \ \ \ W.de Melo and S. van Strien, \textit{One Dimensional Dynamics} ,

\ \ \ \ \ \ \ \ \ Springer- Verlag , Berlin ,1993 .

[98] \ \ M.Hamermesh, \textit{Group Theory}, Addison-Wesley ,

\ \ \ \ \ \ \ \ \ Reading , MA ,1964 .

[99] \ \ \ J.de Azcarraga and J.Izquierdo, \textit{Lie Groups, Lie Algebras,}

\ \ \ \ \ \ \ \ \ \textit{Cohomology and Some Applications in Physics},

\ \ \ \ \ \ \ \ \ Cambridge U.Press,Cambridge,1998 .

[100] \ Y.Choquet-Bruhat and C.DeWitt-Morette,\textit{ Analysis, Manifolds}

\ \ \ \ \ \ \ \ \ \ \textit{and \ Physics}, Part II, Nort-Holland, Amsterdam
,1989 .

[101] \ \ N.Kuiper,Michigan Math.Journ.\textbf{2 (}1952\textbf{)} 95.

[102] \ \ L.Bers,Bull.Am.Math.Soc. (New Series) \ \textbf{5 (}1981\textbf{)}131.

[103] \ \ A.Zamolodchikov, JETP Lett.\textbf{43} (1986) 730;

\ \ \ \ \ \ \ \ \ \ Sov.J.Nucl.Phys.\textbf{46} (1987)1090.

[104] \ \ G.Mostow, I.H.E.S.Publ.\textbf{34} (1968)53.

[105] \ \ \ S.Donaldson and D.Sullivan, Acta Math.\textbf{163 (}1989\textbf{)}181.

[106] \ \ \ S.Matsumoto, Adv.Stud. in Pure Math.\textbf{20 }(1992)167 .

[107] \ \ \ M.Gromov,H.Lawson Jr.and W.Thurston,

\ \ \ \ \ \ \ \ \ \ \ I.H.E.S. Publ.\textbf{68} (1988) 27.

[108] \ \ \ N.Kuiper, I.H.E.S.Publ. \textbf{80} (1988) 47.

[109] \ \ \ H.Reimann, Ann.Acad.Sci.Fenn.\textbf{10}(1985) 477.

[110] \ \ \ M.Kapovich, Ann.Acad.Sci. Fenn. \textbf{23} (1998) 83.

[111] \ \ \ M.Hennington and K.Skanderis, ''The holographic Weil

\ \ \ \ \ \ \ \ \ \ \ \ anomaly'', hep-th/9806087.

[112] \ \ \ D.Johnson and J.Millson, ''Deformation spaces associated

\ \ \ \ \ \ \ \ \ \ \ to compact hyperbolic manifolds'', in \textit{Discrete
Groups in}

\ \ \ \ \ \ \ \ \ \textit{\ \ Geometry and Analysis}, Birh\"{a}user, Boston,1987.

[113] \ \ \ I.Kra , \textit{Automorphic Forms and Kleinian Groups},

\ \ \ \ \ \ \ \ \ \ \ W.A.Benjamin, Reading , MA,1972 .

[114] \ \ \ A.Borel and N.Wallach , \textit{Continuous Cohomology, }

\ \ \ \ \ \ \ \ \ \ \ \textit{Discrete\ Subgroups and Representations of
Reductive }

\ \ \ \ \ \ \ \ \ \ \ \textit{Groups}, Ann.Math.Studies \textbf{94} ,
Princeton U.Press ,1980 .

[115] \ \ \ S.Eilenberg and S.McLane, Ann.Math.\textbf{48} (1947) 51.

[116] \ \ \ C.Kourouniotis, Math.Proc. Cambr.Phil.Soc.\textbf{ 98} (1985) 247.

[117] \ \ \ A.Balinskii and S.Novikov, Sov.Math. Dokl. \textbf{32}(1985) 228.
\end{document}